\PassOptionsToPackage{nobiblatex}{rsproca_new}  % 新增这一行
\documentclass[openacc]{rsproca_new}%%%%where rsproca is the template name

\usepackage{lipsum}
\usepackage{multicol}
\usepackage{amsmath}
\usepackage{graphicx}
\usepackage{float}
\usepackage{placeins}
\usepackage{subcaption}
\usepackage{caption}
\usepackage{csquotes}
\usepackage{textcomp} 
\usepackage{cite}  % 保留这个 BibTeX 引用包
\jname{rsif}
\Journal{J R Soc Interface\ }

\begin{document}

%%%% Article title to be placed here
\title{Cooperation in Public Goods Games over Uniform Random Hypergraphs with Game Transitions}

\author{%%%% Author details
Nankun~Wei$^{1}$, Xiaojin~Xiong$^{1}$, Qin~Li$^{2}$, Minyu~Feng$^{1}$, Attila~Szolnoki$^{3}$}

%%%%%%%%% Insert author address here
\address{$^{1}$College of Artificial Intelligence, Southwest University\\Chongqing 400715, PR China\\
$^{2}$Business College, Southwest University\\Chongqing 402460, PR China\\
$^{3}$Institute of Technical Physics and Materials Science, Centre for Energy Research\\P.O. Box 49, Budapest H-1525, Hungary}

%%%% Subject entries to be placed here %%%%

% \subject{xxxxx, xxxxx, xxxx}

%%%% Keyword entries to be placed here %%%%

%%%% Insert corresponding author and its email address}

\corres{
Qin Li\\
\email{qinli1022@gmail.com}\\
Minyu Feng \\
\email{myfeng@swu.edu.cn}}

%%%% Abstract text to be placed here %%%%%%%%%%%%
\begin{abstract}
The evolution of cooperation is a central enigma in evolutionary game theory. Traditionally, the combination of pairwise networks and repeated Public Goods Games with a single state fails to adequately describe realistic group interaction scenarios. On the one hand, pairwise networks lack clear group definitions. On the other hand, a participant\textquotesingle s decision affects not only competitors' fitness but also the state of the surrounding environment. To address this problem, we propose a Public Goods Game with game transition mechanisms based on Uniform Random Hypergraphs. In our model, game groups formed by hyperedges transition between two types of games, one with abundant public resources and the other with scarce public resources. The transition probability is closely related to the strategies of players within the hyperedges. By developing a Monte Carlo simulation framework that incorporates payoff accumulation, strategy imitation, and game state transitions, we aim to reveal the coevolutionary patterns of strategies and game states in group interactions. Our study highlights a nonlinear relationship between defection sensitivity and cooperation frequency under game transitions, as well as the asymmetric effects of the two sensitivities in state-dependent transitions. These observations open new directions for how to approach social dilemmas.

\end{abstract}
%%%%%%%%%%%%%%%%%%%%%%%%%%%

%%%%%%%%%% Insert the texts which can accomdate on firstpage in the tag "fmtext" %%%%%

\begin{fmtext}

\section{Introduction}
Cooperative behaviors are prevalent in both natural and social systems and have attracted extensive scholarly attention \cite{nowak_11, rand2013human}. From teamwork and resource sharing in human societies to collaborative scheduling of distributed energy nodes in smart grids, cooperative interactions among individuals play an indispensable role. However, when individuals face the dilemma between \enquote{personal gain} and \enquote{collective benefit}, rational choices often favor defection. This makes the emergence and maintenance of group cooperation in social dilemmas a pressing scientific challenge \cite{dawes1980social, nowak2006five}. Evolutionary game theory provides a robust theoretical foundation and an effective framework for studying this phenomenon. Within this framework, game models such as the Prisoner\textquotesingle s Dilemma \cite{yue2025coevolution,zeng2025complex,feng2024information}, Donation Game\cite{zeng2025bursty}, Snowdrift Game \cite{pi2024memory,zeng2022spatial}, and Public Goods Game \cite{szolnoki2016competition, zhang2025evolution} have been extensively studied.
    
\end{fmtext}

\maketitle

\begin{multicols}{2}
As a classic multiplayer game model, the Public Goods Game (PGG) reveals the fundamental logic of social dilemmas: rational defection by participants leads to the tragedy of the commons \cite{hardin1968tragedy}, while mutual cooperation among individuals yields higher collective benefits \cite{milinski2002reputation, kraak2010exploring}.
To investigate the factors influencing cooperative behaviors in Public Goods Games, scholars have thoroughly explored various mechanisms, such as rewards \cite{szolnoki2010reward}, punishments \cite{helbing2010defector}, heterogeneity \cite{perc_pre08, santos_n08}, reputation \cite{he2026reputation}, memory \cite{li2013one, xu2019role, huang2024memory}, and strategy persistence \cite{szolnoki_pre09}. Many of these investigations into mechanisms are conducted under the spatial structure characterized by pairwise networks, because depicting real-world group interactions requires not only well-defined interaction rules but also a clear spatial structure as the underlying carrier. Some studies have systematically summarized the differences in cooperative dynamics of Public Goods Games across classical pairwise networks (including regular lattices, heterogeneous scale-free networks, and small-world networks) \cite{perc2013evolutionary}. Nevertheless, a fundamental limitation of pairwise networks as the carrier for group interactions is their failure to provide a rigorous and unambiguous definition of groups \cite{alvarez2020evolutionary}. In Fig.~S1 of the Supplementary Material, specific examples are used to illustrate that ambiguous group definitions directly lead to a series of issues in game interactions.

Furthermore, the environments of group games in the real world are not static. Consequently, in recent years, the academic community has grown increasingly interested in the coevolutionary dynamics of strategies and environments \cite{hauert2019asymmetric, ito2024complete, tilman2020evolutionary}. When individual strategies not only influence immediate payoffs but also reshape subsequent game types through resource allocation or group dynamics, the so-called \textquotedblleft game transitions\textquotedblright occur \cite{hilbe2018evolution}. Addressing this phenomenon, Su {\it et al.} \cite{su2019evolutionary} found that dynamic switching between game states of varying dilemma strengths can lower the critical threshold for the evolution of cooperation, while Feng {\it et al.} \cite{feng2018evolutionary} demonstrated that Markov-based game state transitions can reinforce cooperation through psychological effects. However, existing studies on multiplayer game transition models under spatial structures are all built on traditional pairwise networks. The ambiguous group definitions of pairwise networks may result in individuals without direct connections being placed in the same group in multiplayer game scenarios, thereby exerting unrealistic impacts on each other\textquotesingle s payoffs.

In recent years, higher-order interaction models, particularly hypergraphs and simplicial complexes, have offered new insights for addressing these challenges \cite{mayfield2017higher, battiston2021physics, zhang2023higher, feng2024hypernetwork}. Hypergraphs directly connect multiple nodes through hyperedges, naturally embodying the group characteristic of all individuals within a group being interconnected, and can provide clear group definitions. They have been applied to scenarios such as Public Goods Games and collective intelligent interactions. Alvarez-Rodriguez {\it et al.} \cite{alvarez2020evolutionary} systematically proposed algorithms for constructing Uniform Random Hypergraphs and heterogeneous hypergraphs, with results indicating that higher-order interactions can significantly enhance cooperation stability. Building on this foundation, some studies \cite{pan2023heterogeneous, zou2024spatial} found that heterogeneous investments effectively promote cooperation across hypergraphs of different orders. Civilini {\it et al.} \cite{civilini2024explosive} explored the differences in collective behavior dynamics between traditional graphs and hypergraphs, revealing that compared with pairwise interactions, higher-order interactions not only foster cooperation in competitive environments but also trigger explosive cooperative behaviors.  Battiston {\it et al.} further emphasized that higher-order structures capture group-level dynamics invisible in pairwise models, shedding light on mechanisms of cooperation and collective behavior in human systems \cite{Battiston2025higher}. Additionally, Shi {\it et al.} \cite{shi2024hypergraph} applied hypergraphs to multi-agent Q-learning dynamics, establishing a theoretical framework based on Uniform Random Hypergraphs. Zou and Huang \cite{zou2025cooperation} also examined the impact of punishment on cooperative dynamics in hypergraphs. These studies demonstrate that hypergraphs provide a more suitable spatial structure for investigating the emergence of cooperation under group interactions, and their clear group definitions ensure that each group is equivalent to a complete subgraph -- perfectly addressing the aforementioned shortcomings of pairwise networks in supporting multiplayer games and game transitions.

By building on these observations, we here propose a public goods game model with transition mechanisms based on Uniform Random Hypergraphs (URH). In particular, we use hyperedges as fundamental game units, with each hyperedge corresponding to a public goods game group. In response to varying environmental conditions, the model distinguishes between high-value and low-value game states through differences in synergy factors. Specifically, high-value games correspond to a higher synergy factor than low-value games, which in turn naturally results in a lower degree of social dilemma for the former. The transition probabilities of hyperedges are determined by the proportion of cooperators within the group, with sensitivity coefficients regulating the responsiveness of different states to defection. This design preserves the advantages of hypergraphs in modeling higher-order interactions while enabling dynamic feedback from collective environments to individual behaviors. Through systematic Monte Carlo simulations, we reveal cooperation dynamics and formation mechanisms in group scenarios, offering new insights into understanding and sustaining cooperation in real-world social dilemmas.

The paper is organized as follows: Section~\ref{sec:model} details the interaction system and specific rules for payoff accumulation, strategy learning, and game transitions. Section~\ref{sec:result} presents the effects of different parameters on system cooperation dynamics, accompanied by analysis. Finally, Section~\ref{sec:conclusion} summarizes the cooperation dynamics observed in the model and outlines future research directions.

\section{Model}
\label{sec:model}
High-order interactions are prevalent in real-world group settings, and individual behaviors shape not only their own and partners\textquotesingle outcomes but also the group environment. For example, overgrazing brings short-term gains to individuals but depletes public resources, undermining long-term benefits. To model how individual actions impact payoffs and the environment, we develop a group interaction system based on Uniform Random Hypergraphs. Here, agents engage in repeated Public Goods Games via hyperedges, updating their strategies while their choices influence the game states of the hyperedges they belong to. This section outlines the model\textquotesingle s core setup: URH construction, strategy update rules, and game transition mechanisms.

\subsection{Interaction Network}
In the framework of hypergraphs, a hypergraph is denoted as \(H(N, L)\), where \(N = \{n_1, n_2, \dots, n_{|N|}\}\) represents a set of \(n\) nodes, each node corresponds to an agent, and \(L = \{l_1, \ldots, l_m\}\) represents a set of \(m\) hyperedges, each hyperedge corresponds to a group where Public Goods Games take place.
Each hyperedge can contain two or more agents, and an agent may belong to multiple different hyperedges simultaneously. Thus, the hyperdegree \(k_i\) is used to characterize the number of hyperedges to which agent \(i\) belongs.

Our model focuses on the overall dynamics exhibited by the system through agent interactions. We assume that each agent has equal status and is indistinguishable from one another, i.e., our interaction system is homogeneous and uniform. Therefore, we use URH \cite{alvarez2020evolutionary} for characterizing the interaction relationships between agents in the system.

For a URH with \(N\) nodes, each node is assigned to a hyperedge with equal probability, and each hyperedge contains exactly \(g\) nodes. Such a uniform hypergraph is termed a \(g\)-order uniform hypergraph, and there exists a critical threshold for the number of hyperlinks \(L_c = (N / g) \ln N\) \cite{alvarez2020evolutionary} to ensure that the hypergraph is fully connected.

\subsection{Strategy Update}
Based on the agent interaction system characterized by Uniform Random Hypergraphs, we allow agents to participate in Public Goods Games with hyperedges as groups. Let \(s_i\) denote the strategy adopted by agent \(i\), which can take two values: 0 represents defection (D), and 1 represents cooperation (C). Here, we take the payoff calculation of agent \(i\) as an example to demonstrate the specific payoff calculation process and strategy update rules.

A random agent \(i\) and one of its hyperedges \(l\) are selected, and all agents in \(l\) participate in Public Goods Games across all their respective hyperedges and accumulate their payoffs. First, we define the payoff of node \(i\) in hyperedge \(l\) as \(\pi_{i, l}\), which is calculated as 
\begin{equation}
\begin{split}
\pi_{i, l} &= r_{l} \cdot \sum_{u \in l} c \cdot s_u - c \cdot s_i,
\end{split}
\label{在一条超边中的收益}
\end{equation}
 where the cooperation cost \(c\) is fixed at 1, and \(\sum_{u \in l} c \cdot s_u\) is the total contribution of cooperators in \(l\). Here, \(r_{l}=R_{l}/g\) is the normalized synergy factor, with \(R_{l}\) the raw synergy factor and \(g\) the size of hyperedge \(l\). The total contribution is multiplied by \(r_{l}\), and agent \(i\) pays the cooperation cost \(c\) according to its current strategy.

We define the final payoff of agent \(i\) as
\begin{equation}
\pi_i = \frac{1}{k_i} \sum_{l' \in \Omega_i} \pi_{i,l'}, 
\label{所有超边中总收益}
\end{equation}
where \(\Omega_i\) denotes the set of all hyperedges containing agent \(i\). Agent \(i\) is required to participate in games and accumulate payoffs across these hyperedges. \(k_i\) represents the hyperdegree of agent \(i\), i.e., the number of hyperedges it belongs to, and is used to average the total payoff.

By analogy, after all agents in \(l\) accumulate their payoffs, the highest earner \(j\) in hyperedge \(l\) is identified, and agent \(i\) imitates the strategy of \(j\) with probability $W_{s_{i} \rightarrow s_j}$ defined as
\begin{equation}
W_{s_{i} \rightarrow s_j} = \frac{1}{\Delta} \left[ \pi_j - \pi_{i} \right]\,.
\label{策略更新}
\end{equation}
This linear imitation mechanism is a commonly used strong selection rule in higher-order interaction game studies \cite{pan2023heterogeneous,alvarez2020evolutionary}. In the formula, $\pi_j - \pi_i$ represents the payoff difference between the two agents. Since $j$ is the agent with the highest payoff in hyperedge $l$, this payoff difference is non-negative. $\Delta$ represents the absolute value of the maximum possible payoff difference in the current system, which is used to ensure the normalization of the imitation probability. We use superscripts for differentiation: \(\pi_D^+\) denotes the maximum possible payoff for defectors, and \(\pi_D^-\) refers to the minimum possible payoff. The notation for cooperators adheres to the same convention.
Thus, the maximum and minimum payoffs for different strategies are calculated as
\begin{equation}
\begin{cases}
\pi_D^+ = r_1 (g - 1), \\
\pi_D^- = 0 ,\\
\pi_C^+ = r_1g - 1 ,\\
\pi_C^- = r_2 - 1,
\end{cases}
\end{equation}
where \(r_1\) and \(r_2\) represent two values of the normalized synergy factor, and their specific meanings are elaborated in Section \ref{sec:game-transition}. From this, $\Delta$ has four possible value combinations. Given $r_2 \leq r_1$, the calculation formula of $\Delta$ under different value ranges of $r_1$ and $r_2$ is derived by comparison:
\begin{equation}
\Delta = 
\begin{cases} 
r_1 (g - 1) - (r_2 - 1), & 0 < r_2 < r_1 < 1 ,\\
r_1 g - r_2, & 0 < r_2 \leq 1 \leq r_1 ,\\
r_1 g - 1, & 1 < r_2 < r_1  .
\end{cases} 
\label{归一化delta}
\end{equation}

\begin{figure*}
\centering
\includegraphics[width=0.95\textwidth]{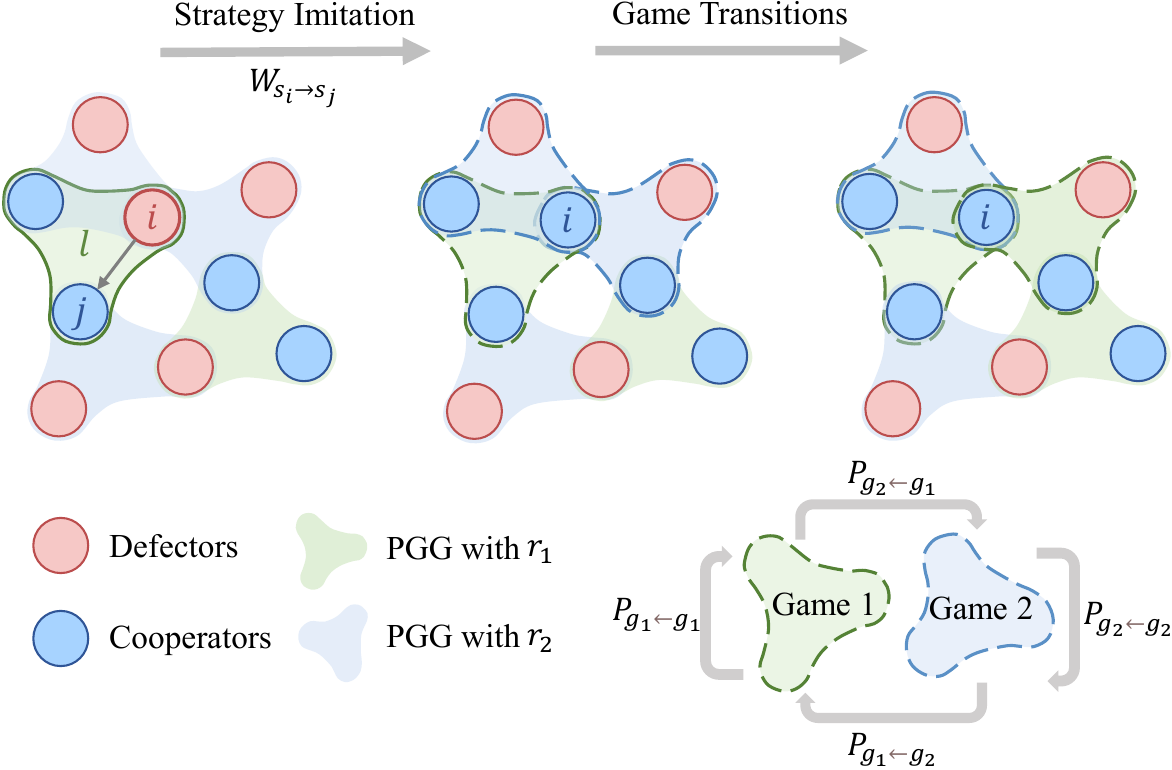} % Adjust width parameter to resize the image
\caption{
    \textbf{Strategy update and game transition process.} 
    Nodes represent agents. Blue nodes are cooperators and red nodes are defectors. Hyperedge colors correspond to game states. Green hyperedges indicate high-value game \( g_1 \) with \( r_1 \) and blue hyperedges indicate low-value game \( g_2 \) with \( r_2 \), and \( r_1 > r_2 \). The dynamics contain two steps. \textbf{(a) Strategy Imitation:} An agent \( i \) is selected randomly and one of its hyperedges \( l \) (solid border) is chosen randomly. Agents in \( l \) accumulate payoffs based on the current game state. Agent \( i \) then imitates the strategy of the highest-earning agent \( j \) in \( l \) with probability \( W_{s_i \rightarrow s_j} \). \textbf{(b) Game Transition:} Following the strategy update step for player \( i \), regardless of whether its strategy is changed or not, all hyperedges containing \( i \) (dashed borders) undergo transitions. These transitions rely on their cooperator fractions and follow the rules at the bottom right. Cooperation promotes transitions to \( g_1 \) while defection leads to \( g_2 \). These two transitions alter payoff structures and produce distinct cooperation dynamics compared to single Public Goods Games.
}
    \label{fig:evolution_process} % Image caption
% Label for cross-referencing
\end{figure*}
% Example of figure reference: See Fig.\ref{fig:model}

To facilitate understanding of the strategy imitation process, we present a partial network schematic in Fig.~\ref{fig:evolution_process}. Agents are represented as nodes, with their strategies distinguished by different colors. The strategy imitation process is probabilistic. In the strategy imitation stage, we take defector agent $i$ as an example, demonstrating a typical scenario where it successfully learns the strategy of agent $j$, the highest earner in the hyperedge, and ultimately switches to cooperation.

\subsection{Game Transition}
\label{sec:game-transition} 
The core idea of game transitions is that players' strategies not only affect current payoffs but also influence subsequent game types. A classic and realistic assumption is that cooperative behaviors improve the game environment, while defection deteriorates it. Therefore, cooperative behaviors increase the probability of transitioning to the high-value game system, while defection increases the probability of transitioning to the low-value system. 
%改

To avoid conceptual confusion, it is necessary to clarify the distinction between our game transition mechanism and reputation mechanism (e.g., the LR2 reputation mechanism \cite{ren2025bottom}). Reputation focuses on interactions and evaluations among individuals, whereas the game transition mechanism centers on the reciprocal influence between individual strategies and the environment. These two mechanisms target different objects of interaction but are not mutually conflicting.

Our research focus is not on designing or calculating transition probabilities between multiple game states. Instead, we prioritize the cooperative dynamics exhibited by the coevolution of strategies and the environment in higher-order spatial structures, hence adopting a two-state game transition framework. Considering the structural characteristics of hypergraphs, we assume each hyperedge \(l\) possesses a unique synergy factor that takes only two possible values: \(r_1\) or \(r_2\) (with \(r_2 \leq r_1\)). Specifically, \(r_1\) corresponds to a state of abundant environmental resources, where a relatively mild social dilemma exists, and players can obtain significant payoffs through cooperation, thus corresponding to high-value games. In contrast, \(r_2\) corresponds to a state of scarce environmental resources, where the efficiency of public resource enhancement is low. Even if individuals choose to cooperate, they struggle to break through the payoff bottleneck due to the more severe social dilemma, making such games low-value ones. To intuitively reflect the resource advantage of the high-value game relative to the low-value game and simplify subsequent descriptions, we define \(\delta = r_1 - r_2\).

Based on this, all interaction groups in the system are divided into two subsystems: the high-value game subsystem \(g_1\) and the low-value game subsystem \(g_2\). Under the transition mechanism, corresponding state transition probabilities exist between these two subsystems.
    
Taking hyperedge \(l\) as an example, Eqs.~\ref{高转入高} and \ref{低转入高} show the probabilities of transitioning into the high-value game when hyperedge \(l\) is in different game states, while \(P_{g_2 \leftarrow g_1} = 1 - P_{g_1 \leftarrow g_1}\) and \(P_{g_2 \leftarrow g_2} = 1 - P_{g_1 \leftarrow g_2}\) represent the probabilities of transitioning into the low-value game, including
\begin{equation}
P_{g_1 \leftarrow g_1} = \left( \frac{\sum_{u \in l} s_u}{g} \right)^{\alpha_1},
\label{高转入高}
\end{equation}
and
% Second equation
\begin{equation}
P_{g_1 \leftarrow g_2} = \left( \frac{\sum_{u \in l} s_u}{g} \right)^{\alpha_2}.
\label{低转入高}
\end{equation}
Here, \(\frac{\sum_{u \in l} s_u}{g}\) denotes the proportion of cooperators within hyperedge $l$, with the basic transition probability directly determined by this proportion. This setting conforms to the classical framework of coevolution between the game environment and agents' strategies in game transitions \cite{hilbe2018evolution,su2019evolutionary}, where cooperation enhances the game environment while defection degrades it. Building on this proportion of cooperators, we incorporate exponential sensitivity coefficients \(\alpha_1\) and \(\alpha_2\) to adjust the basic probability. This design is widely utilized in coevolutionary mechanism research \cite{pan2023heterogeneous,szolnoki_pre09} as exponential sensitivity avoids the uniform environmental response to strategy changes inherent in linear functions, enabling the characterization of varied environmental sensitivity to defection. Specifically, \(\alpha_1\) and \(\alpha_2\) regulate the sensitivity of hyperedges to defection across different subsystems and define the nature of game transitions. In the context of subsystem transitions, a transition is state-independent if its probability depends exclusively on the target subsystem\textquotesingle s characteristics rather than the current subsystem. Conversely, it is state-dependent if the probability of transitioning to the same target subsystem differs based on the current subsystem.

%改

First, we consider the case where $\alpha_1 = \alpha_2$. In this case, the state transition has no memory effect. The transition probability is independent of the current and past states of the hyperedge, i.e., the game transition is state-independent. To simplify, we denote \(\alpha_1 = \alpha_2 = \alpha\). When \(\alpha = 0\), the entire system degenerates back to a single public goods game of pure \(g_1\), and the game type of the hyperedge does not respond to the players' defection behaviors. Suppose that the value of \(\alpha\) gradually increases. In that case, it means that the environment in the system becomes more sensitive to the defection of agents, and a hyperedge can only transition into the high-value game system with a high probability when the number of defectors in it is very small.

\begin{figure*}
\centering
% First row of subfigures
\begin{subfigure}[b]{0.30\textwidth}
\centering
\includegraphics[width=\textwidth]{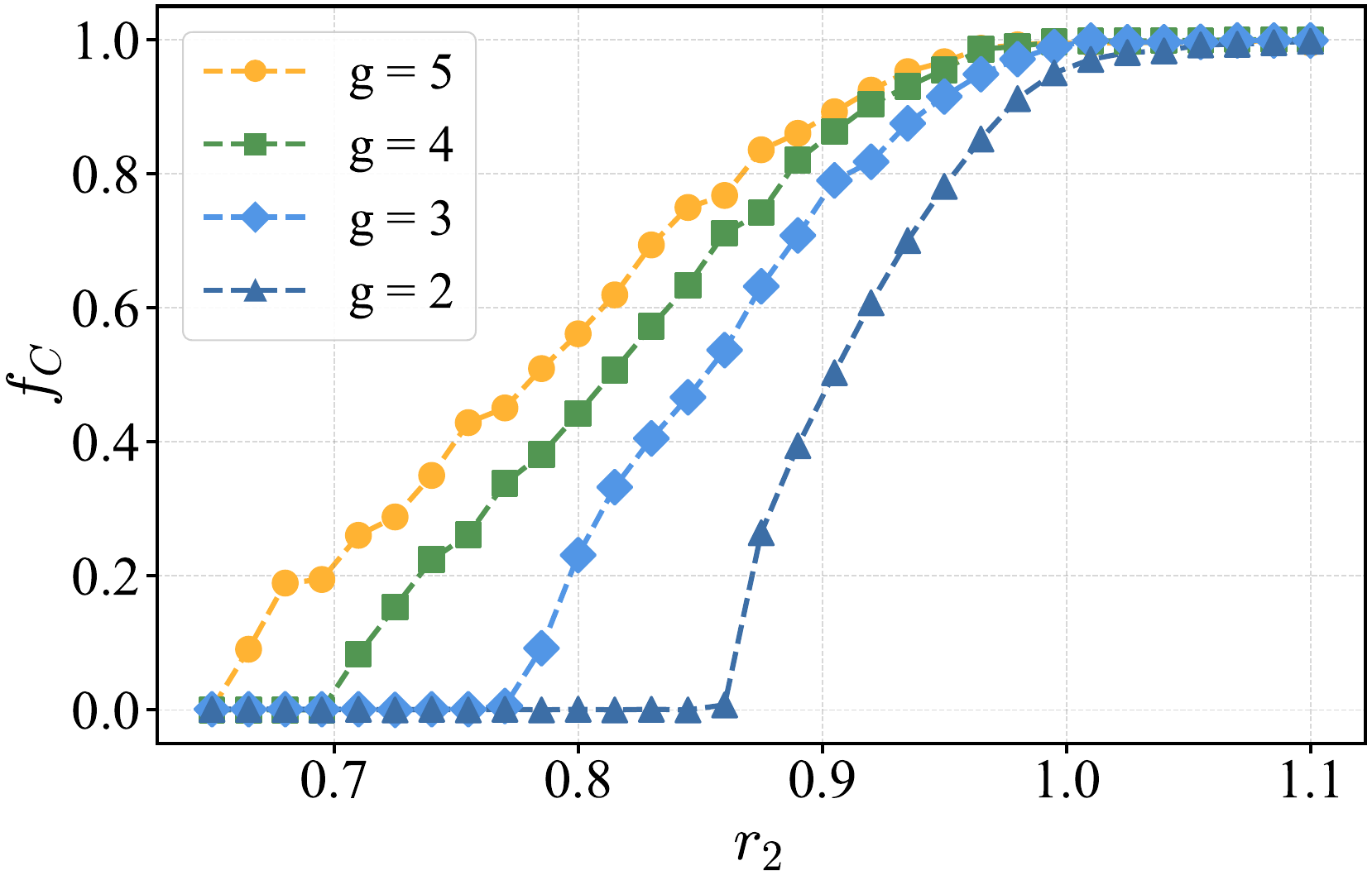}
\caption{\(\alpha=0.00,\delta=0.05\)}
\label{fig:exp1_single_high}
\end{subfigure}
\hfill
\begin{subfigure}[b]{0.30\textwidth}
\centering
\includegraphics[width=\textwidth]{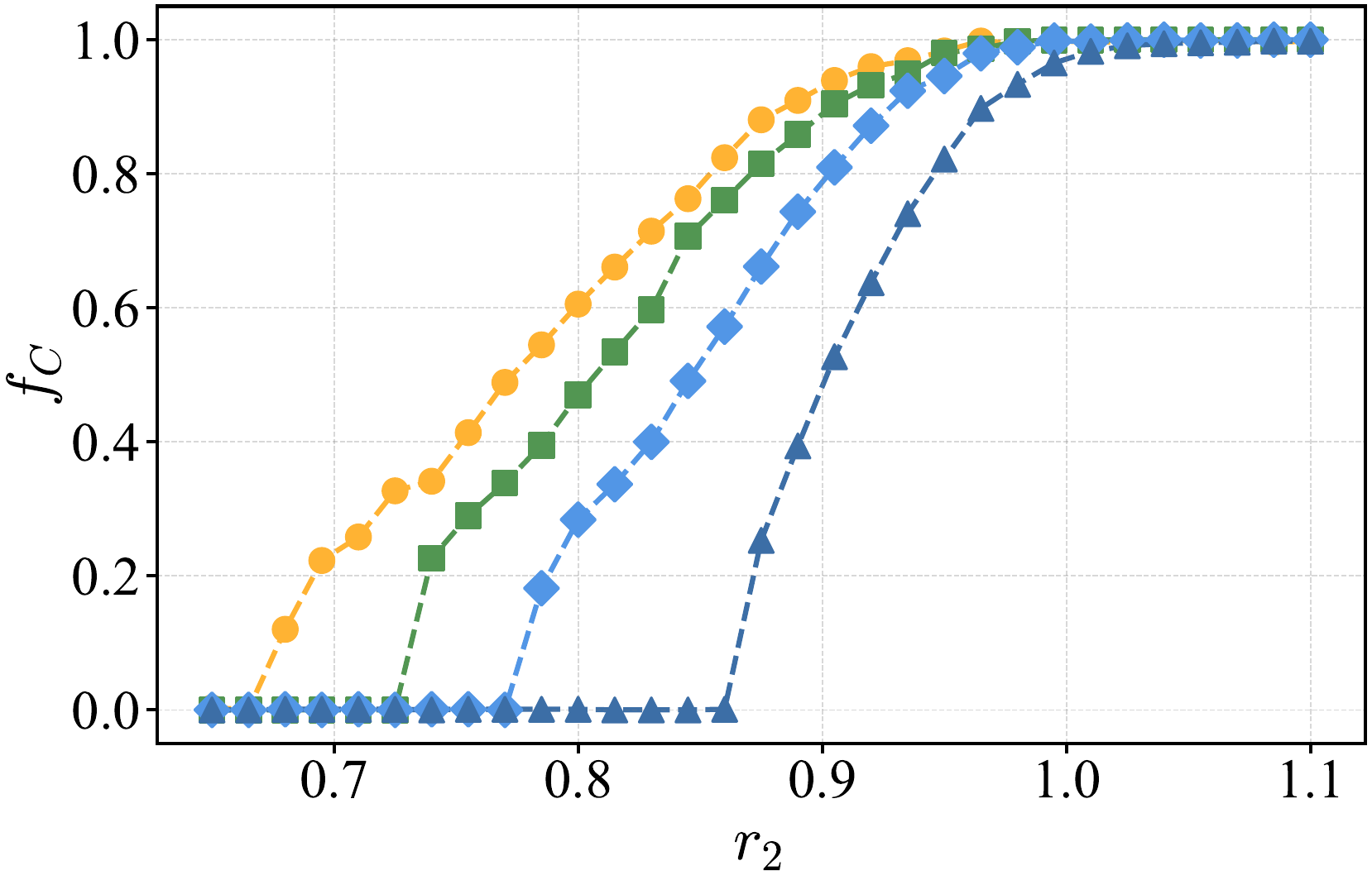}
\caption{\(\alpha=0.50,\delta=0.05\)}
\label{fig:exp1_alpha_0.5}
\end{subfigure}
\hfill
\begin{subfigure}[b]{0.30\textwidth}
\centering
\includegraphics[width=\textwidth]{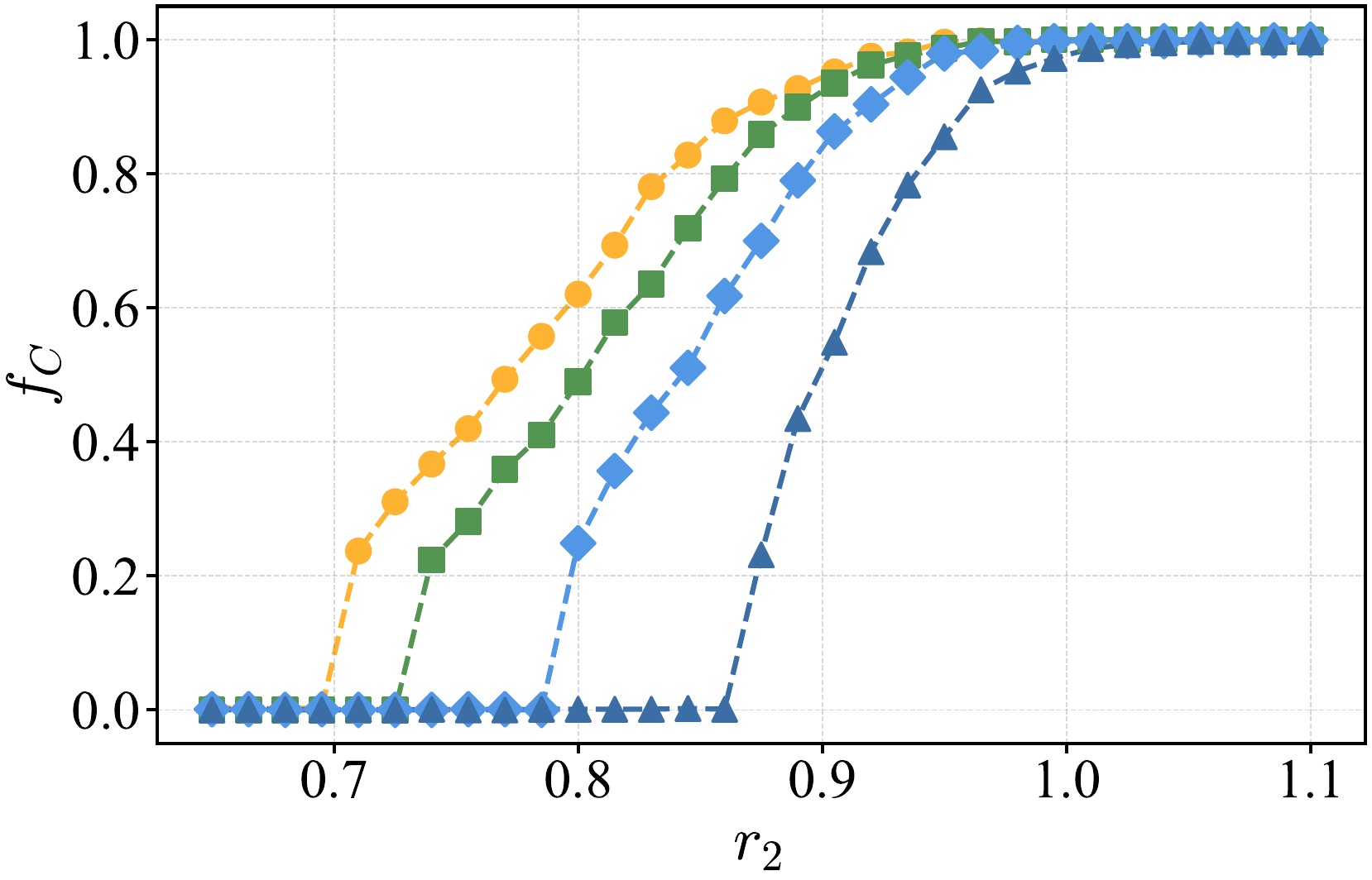}
\caption{\(\alpha=1.00,\delta=0.05\)}
\end{subfigure}

% Second row of subfigures
\vspace{0.1cm} % Vertical spacing between rows
\begin{subfigure}[b]{0.30\textwidth}
\centering
\includegraphics[width=\textwidth]{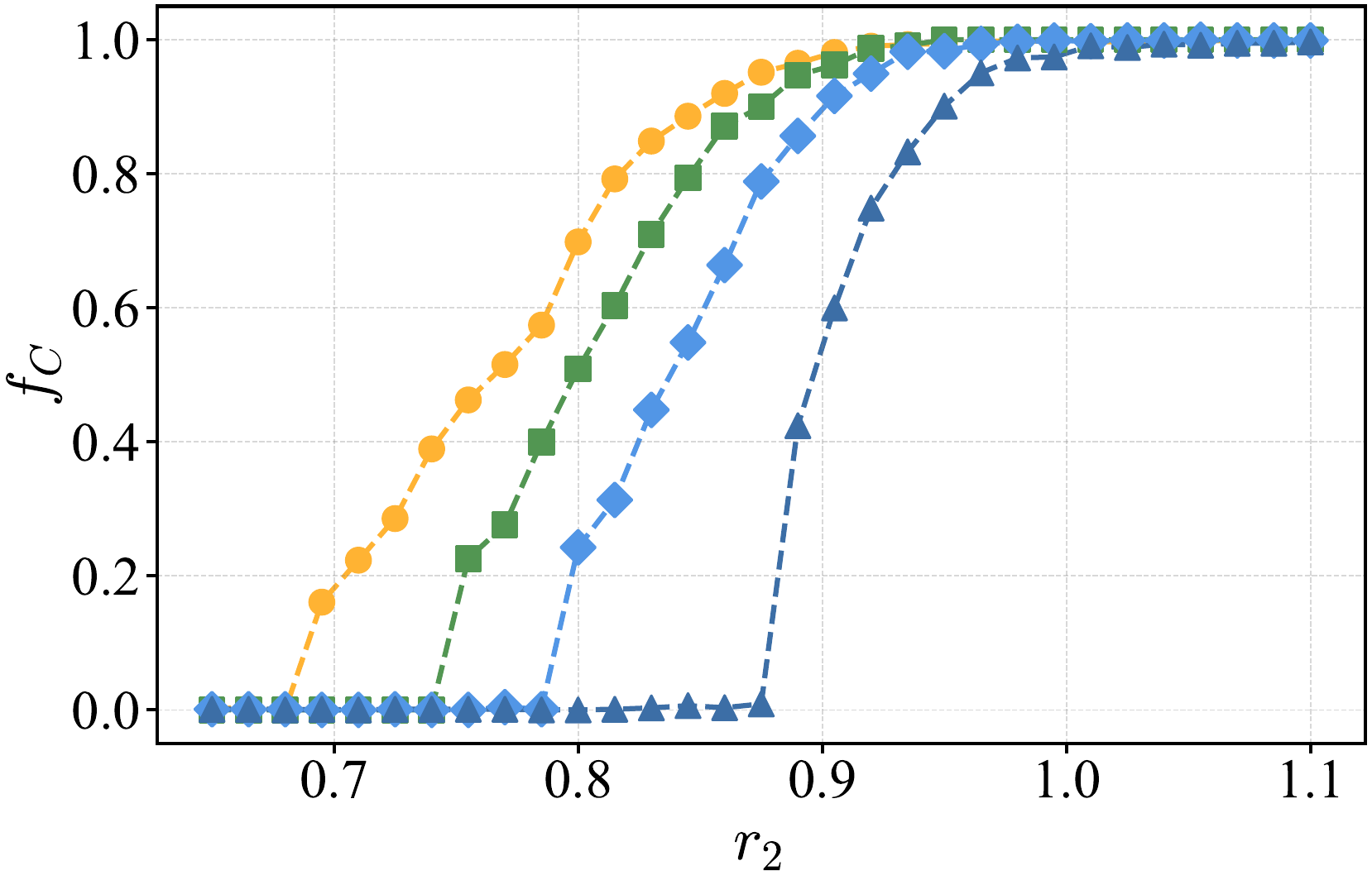}
\caption{\(\alpha=2.00,\delta=0.05\)}
\end{subfigure}
\hfill
\begin{subfigure}[b]{0.30\textwidth}
\centering
\includegraphics[width=\textwidth]{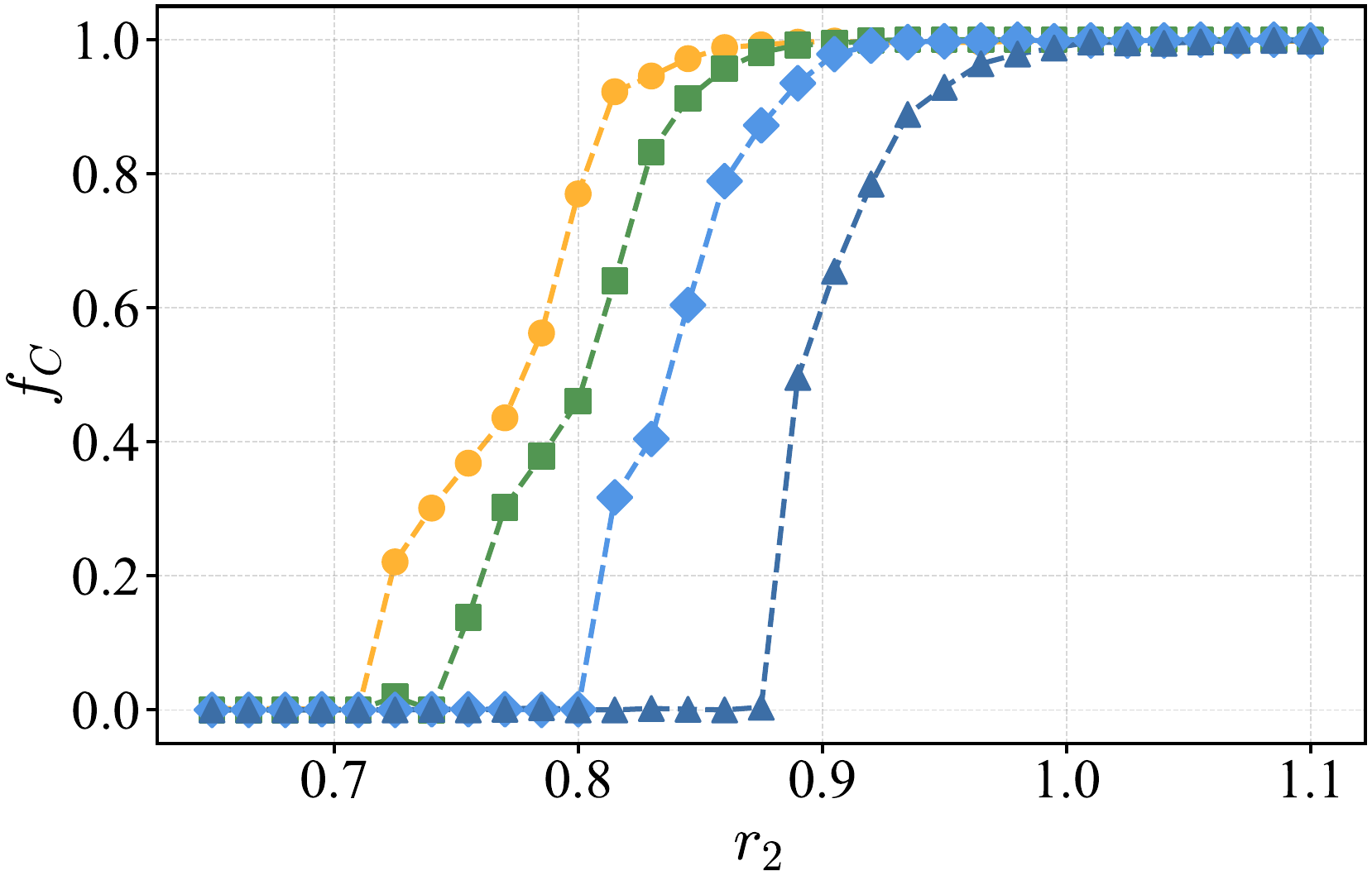}
\caption{\(\alpha=5.00,\delta=0.05\)}
\label{fig:exp1_alpha_5}
\end{subfigure}
\hfill
\begin{subfigure}[b]{0.30\textwidth}
\centering
\includegraphics[width=\textwidth]{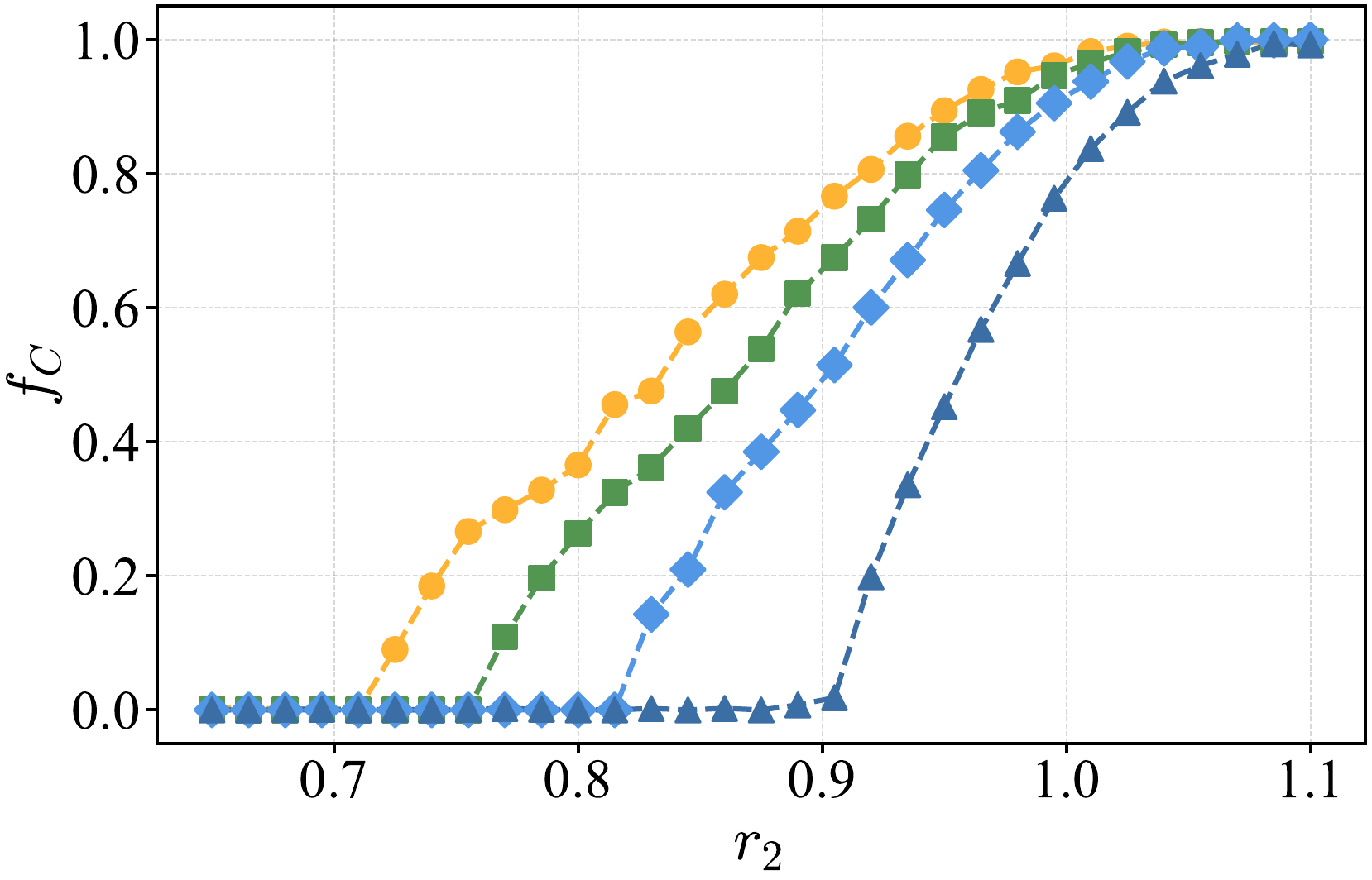}
\caption{\(\delta=0.00\)}
\label{fig:exp1_single_low}
\end{subfigure}
\caption{\textbf{Cooperation frequency \(f_C\) as a function of \(r_2\) under different sensitivity levels.} The x-axis represents the low-value synergy factor \(r_2\), with curves for different hypergraph orders: yellow circles (\(g=5\)), green squares (\(g=4\)), blue diamonds (\(g=3\)), and purple triangles (\(g=2\)). Panels~(\subref{fig:exp1_single_high})-(\subref{fig:exp1_alpha_5}) set \(\delta=0.05\) with \(\alpha=0, 0.5, 1, 2, 5\), respectively. Panel~(\subref{fig:exp1_single_low}) shows the single low-value game without game transitions (\(\delta=0\)). From Figs.~(\subref{fig:exp1_single_high})-(\subref{fig:exp1_alpha_5}), as the system\textquotesingle s sensitivity to defection increases, this transition mechanism raises the cooperation threshold while amplifying \(f_C\) growth to boost cooperation for high \(r_2\). This compresses the \(r_2\) interval required for \(f_C\) to rise from 0 to 1, leading to a steeper curve trend.}
\label{fig:exp1}
\end{figure*}
\begin{figure*}
\centering
% First row
\begin{subfigure}[b]{0.32\textwidth}
\centering
\includegraphics[width=\textwidth]{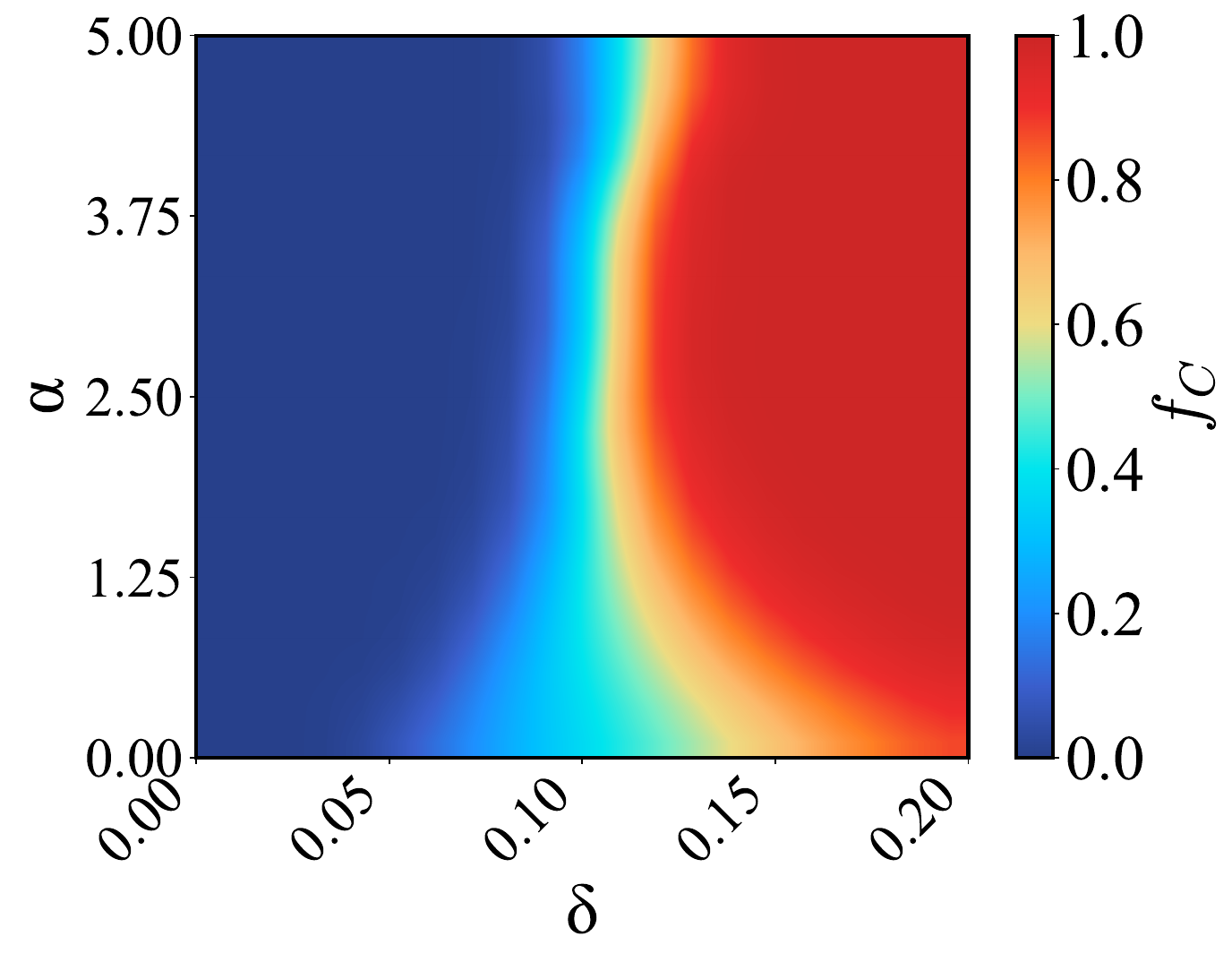}
\caption{\(g=3\), \(r_2=0.775\)}
\label{fig:exp2_g3_collapse}
\end{subfigure}
\hfill
\begin{subfigure}[b]{0.32\textwidth}
\centering
\includegraphics[width=\textwidth]{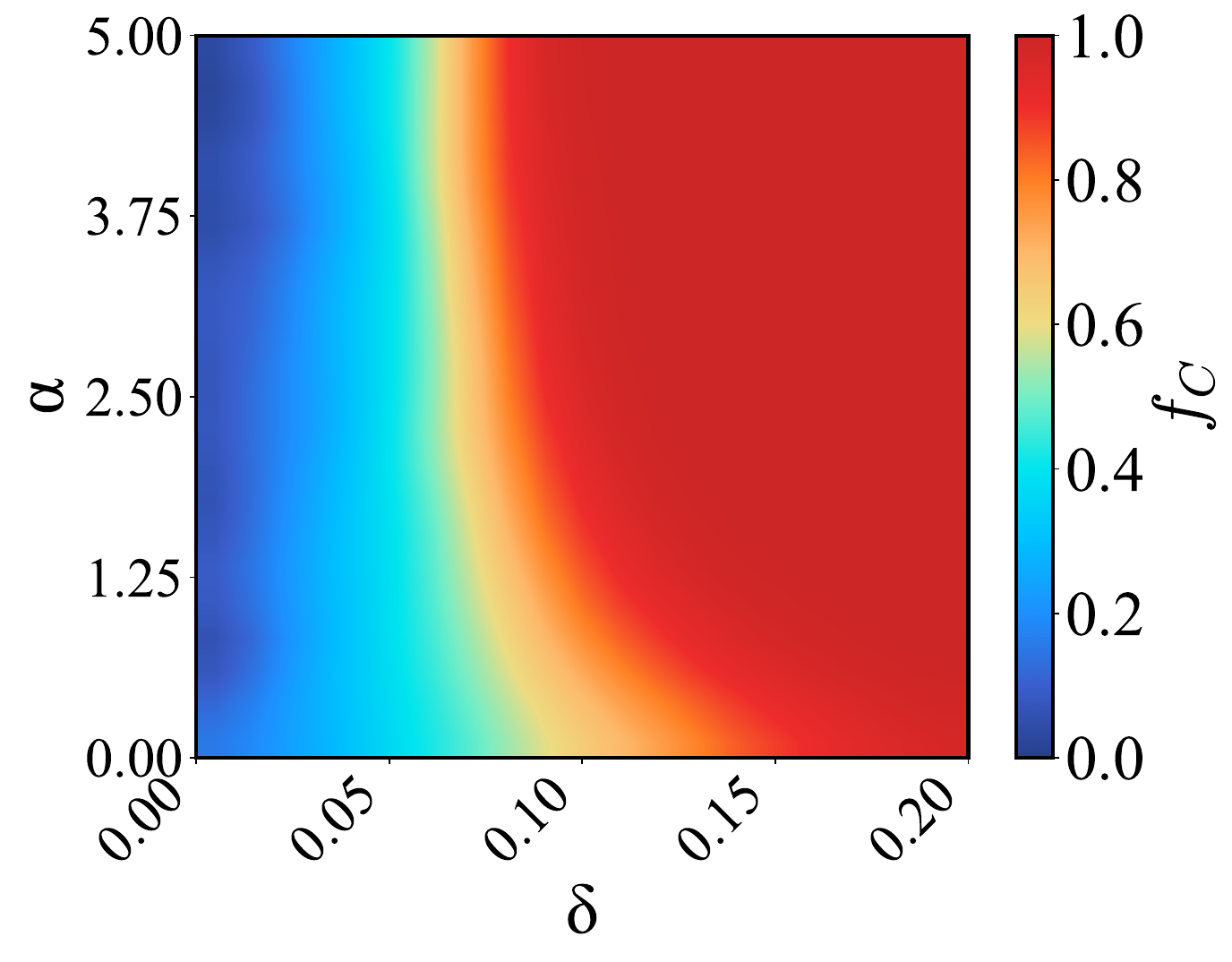}
\caption{\(g=3\), \(r_2=0.825\)}
\label{fig:exp2_g3_critical}
\end{subfigure}
\hfill
\begin{subfigure}[b]{0.32\textwidth}
\centering
\includegraphics[width=\textwidth]{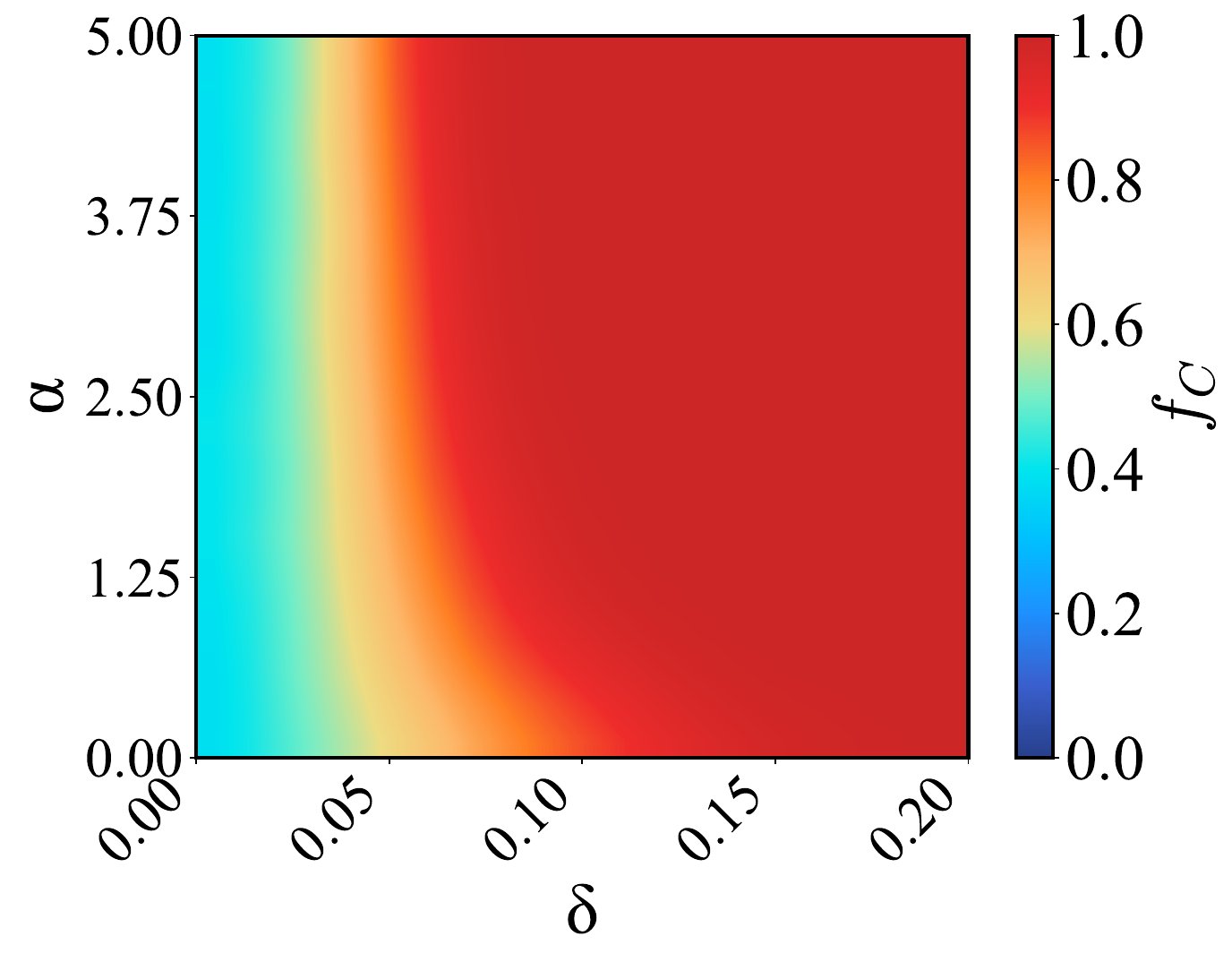}
\caption{\(g=3\), \(r_2=0.875\)}
\label{fig:exp2_g3_sustain}
\end{subfigure}

% Second row
\vspace{0.1cm}
\begin{subfigure}[b]{0.32\textwidth}
\centering
\includegraphics[width=\textwidth]{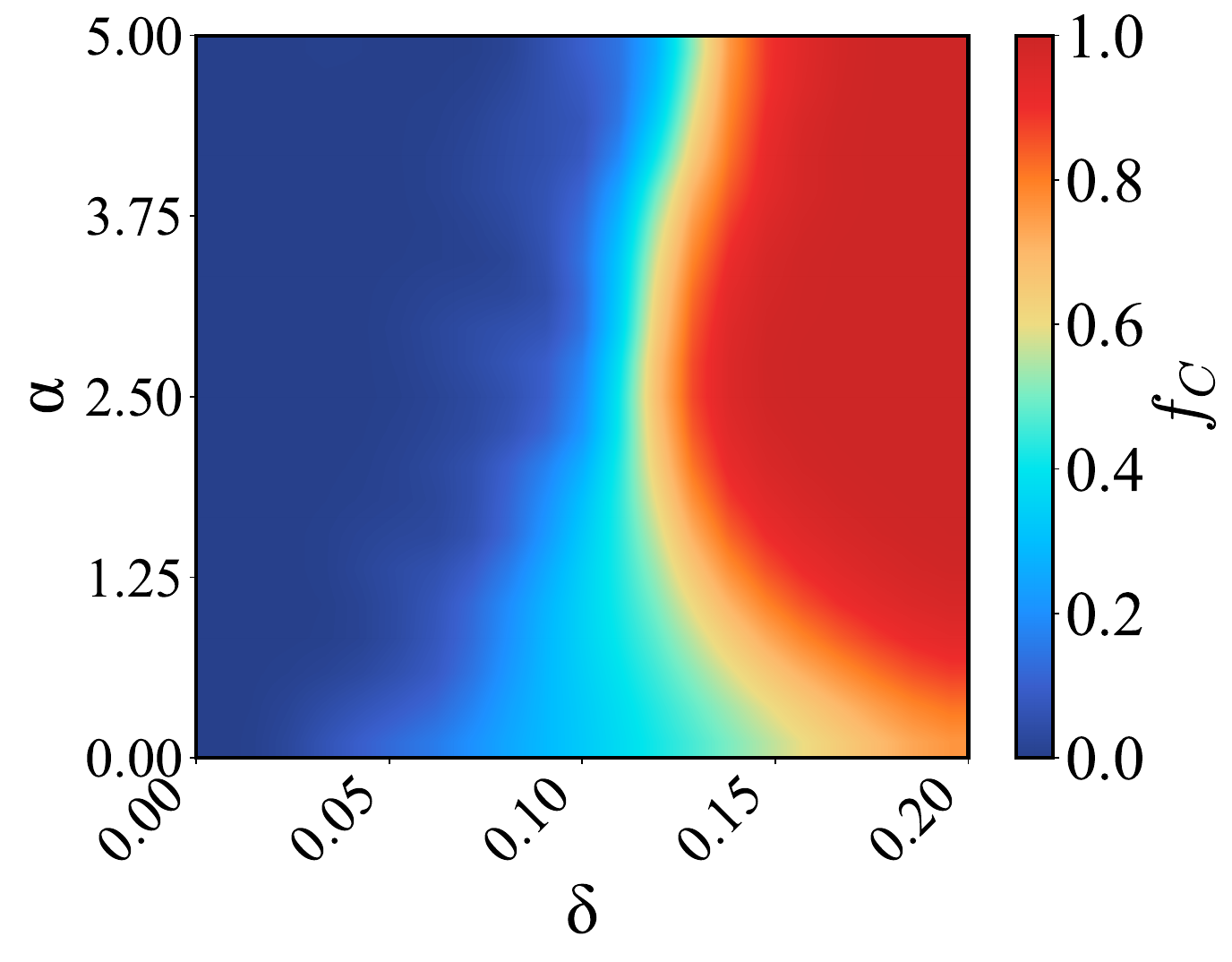}
\caption{\(g=4\), \(r_2=0.720\)}
\label{fig:exp2_g4_collapse}
\end{subfigure}
\hfill
\begin{subfigure}[b]{0.32\textwidth}
\centering
\includegraphics[width=\textwidth]{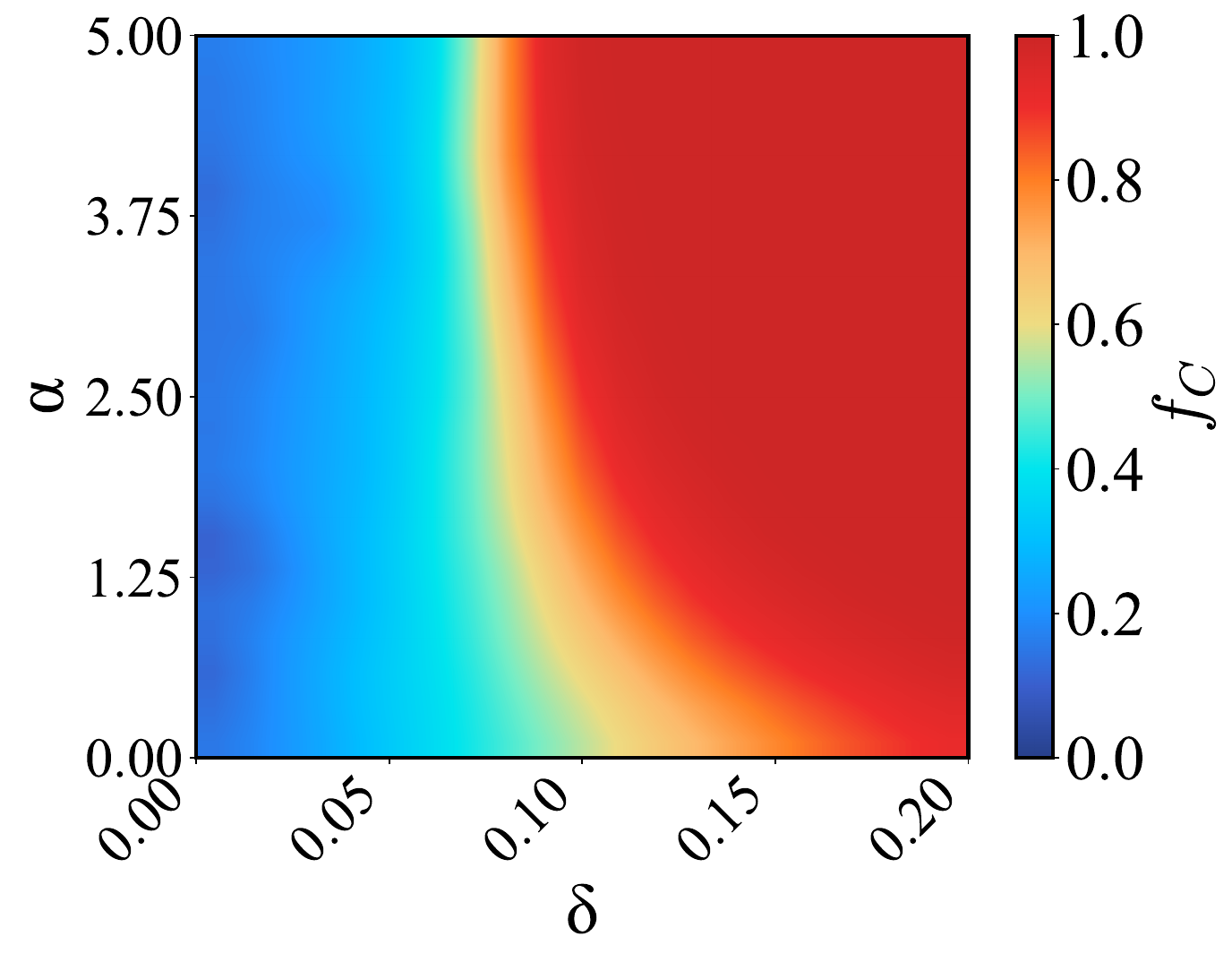}
\caption{\(g=4\), \(r_2=0.770\)}
\label{fig:exp2_g4_critical}
\end{subfigure}
\hfill
\begin{subfigure}[b]{0.32\textwidth}
\centering
\includegraphics[width=\textwidth]{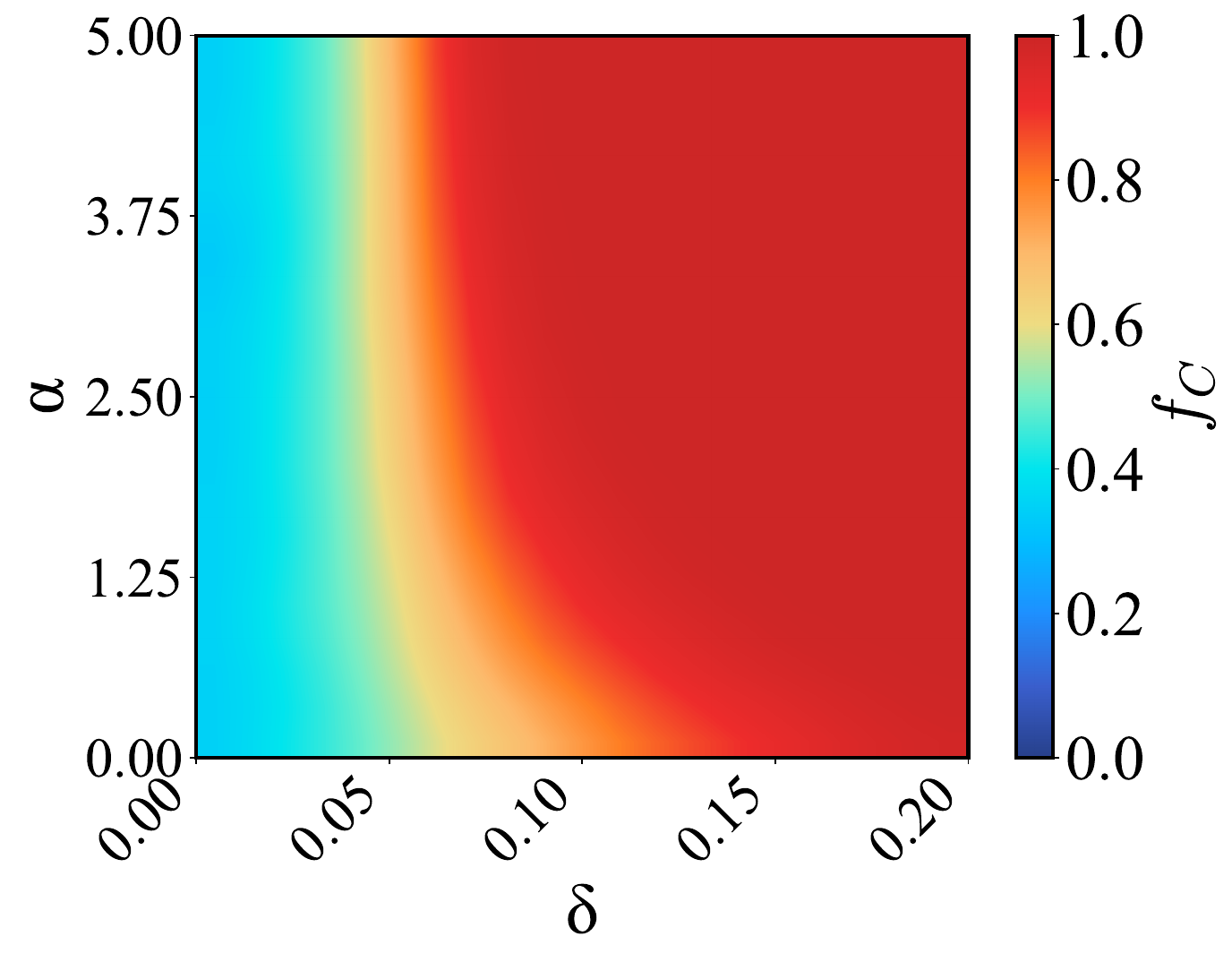}
\caption{\(g=4\), \(r_2=0.820\)}
\label{fig:exp2_g4_sustain}
\end{subfigure}

\caption{\textbf{Heatmaps of cooperation frequency \(f_C\) as functions of \(\delta\) and \(\alpha\) for three representative \(r_2\) values.} The first row corresponds to \(g=3\) with \(r_2=0.775, 0.825, 0.875\). The second row corresponds to \(g=4\) with \(r_2=0.72, 0.77, 0.82\). Heatmaps are mildly smoothed with a Gaussian filter (\(\sigma=1\)) to suppress numerical fluctuations, and bilinear interpolation is used for visualization. At the boundaries between pure cooperation and pure defection for different hypergraph orders, three consistent phenomena are observed as the sensitivity \(\alpha\) increases: cooperation is inhibited, the cooperation level remains stable, and cooperation is promoted.}
\label{fig:exp2}
\end{figure*}

%改
Next, we consider the case where $\alpha_1 \neq \alpha_2$. In this case, the game transition is state-dependent. Such transitions satisfy the Markov property, the transition probability at the next step depends only on the current state,
and is independent of the past states of the hyperedge \cite{hilbe2018evolution}. When \(\alpha_1 < \alpha_2\), the hyperedges in \(g_1\) are less sensitive to defection than those in \(g_2\). If two hyperedges have the same number of cooperators, the hyperedge in the \(g_1\) state is more likely to remain in \(g_1\) in the future, while the hyperedge in the \(g_2\) state is more difficult to transition into \(g_1\), vice versa when \(\alpha_1 > \alpha_2\).

In Fig.~\ref{fig:evolution_process}, we continue to focus on agent $i$ and explain the impact of its strategy change on the hyperedges it belongs to. Note that only one possible transition outcome is shown in the figure, so the state dependence of transition probabilities is not reflected. Specifically, the transition results of the three hyperedges marked by dashed lines in the figure are as follows: the hyperedge composed entirely of cooperators will remain in the high-value game state with a probability of 1. Among the other two hyperedges, one stays in the low-value game state represented by blue, and the other successfully transitions to the high-value game state and turns green.

\section{Simulation and Analysis}
\label{sec:result}
This section investigates the cooperation dynamics of Public Goods Games with transition mechanisms on Uniform Random Hypergraphs (URH) through Monte Carlo simulations. We analyze the system by extending game transitions from state-independent to state-dependent scenarios, and employ asynchronous random sequential updating.

The system is uniformly initialized with equal proportions. Agents are randomly assigned to either cooperation or defection, each accounting for 50\%. Similarly, hyperedges are randomly initialized to either high-value ($g_1$) or low-value ($g_2$) game states, each accounting for 50\%. To ensure the reliability of the results, we verify the robustness of the data with different initialization ratios of cooperators and hyperedge states, and detailed analyses are provided in Section S2 of the Supplementary Material.

One Monte Carlo step consists of $N$ elementary steps. In each elementary step, a target node and one of its affiliated hyperedges are selected. Subsequently, payoffs for all players in the target hyperedge are accumulated, the target node\textquotesingle s strategy is updated, and game transitions are performed for all hyperedges containing the target node. To ensure stable results, we construct an URH with $|N| = 1000$ and $|L| = L_c$, set the iteration steps to $T = 10^4$, and use data from the last 4000 steps for averaging under different parameter combinations. To ensure the stability of the numerical results, we calculated the average of the critical value data points using data from at least 30 independent simulations.

\subsection{State-Independent Transitions}
Unlike traditional single-game scenarios, the transition mechanism enables hyperedges to switch between high-value (\(g_1\)) and low-value (\(g_2\)) game states. A natural question is whether the overall dynamics exhibited by the system represent a simple superposition of the cooperative dynamics of these two individual game states.

To verify this conjecture, we first set \(\delta = r_1 - r_2 = 0.05\) to ensure a moderate difference between the two game states. We then examine how the cooperation frequency \(f_C\) varies with the low-value synergy factor \(r_2\) under different sensitivity levels \(\alpha\), and the results are shown in Fig.~\ref{fig:exp1}.

Fig.~\ref{fig:exp1} first illustrates how the cooperation frequency associated with the sensitivity coefficient \(\alpha\) behaves across the pairwise and higher-order interactions characterized by the URH, under the state-independent condition \(\alpha_1 = \alpha_2 = \alpha\). We define the critical value \(r_{2c}\) as the minimum \(r_2\) required to sustain cooperation. Overall, as the interaction order $g$ increases, the critical \(r_{2c}\) for higher-order interactions (\(g > 2\)) to maintain cooperation is significantly lower than that for pairwise interactions (\(g = 2\)). This result aligns with findings reported in \cite{civilini2024explosive,alvarez2020evolutionary}.

The sensitivity coefficient $\alpha$ quantifies a hyperedge\textquotesingle s sensitivity to defection, and its regulatory effect on cooperation shows a nonlinear trend with the increase of $r_2$. Specifically, the cooperation extinction threshold $r_{2c}$, a higher $\alpha$ elevates the $r_{2c}$ threshold more markedly, indicating that increased defection sensitivity suppresses cooperation under strong social dilemmas in $g_2$.

In contrast, as $r_2$ increases, that is, as the social dilemma intensity in the low-value game $g_2$ weakens, the cooperation-promoting effect of the transition mechanism gradually emerges. A higher sensitivity coefficient $\alpha$ significantly boosts the cooperation frequency, and this promoting effect is more pronounced in higher-order interactions.
% For instance, in the case of the yellow curve corresponding to the interaction order $g=5$, increasing $\alpha$ from 0 to 5 raises the cooperation frequency from 0.56 to 0.78 when $r_2=0.8$. In comparison, the promoting effect of $\alpha$ on pairwise interactions with the interaction order $g=2$ is much weaker.

This is because pairwise interactions involve small groups and have low tolerance for defection. Even if the social dilemma is alleviated, a group can switch to the high-value game only when both individuals cooperate. In contrast, higher-order interactions tolerate defection better. For example, when a group has only one defector, a five-player group ($g=5$) has a much higher probability of switching to the high-value game than a pairwise group.
%改
To verify the universality of the URH, we additionally conduct comparative simulations on the \textit{Erd\H{o}s--R\'enyi} (ER) random network to characterize pure pairwise interaction scenarios. The results (Fig. S2 of the Supplementary Material) demonstrate that the URH can not only accommodate the characterization of higher-order interactions through flexible adjustment of the interaction order \(g\) but also accurately describe pairwise interactions when \(g = 2\). 
In addition, to validate the robustness of our conclusions, we conduct supplementary simulations on randomly connected mixed-order hypergraphs (see Section S3 in the Supplementary Material for details). Results confirm that our core findings remain consistent under mixed-order structures: cooperation is still significantly promoted as the average interaction order increases.

These further support the core conclusion on the URH. Higher-order interactions promote cooperation more effectively than pairwise interactions. Under the transition mechanism, the nonlinear influence of the defection sensitivity coefficient $\alpha$ on $f_C$ is more significant in higher-order interactions.

%改
With the variation of $r_2$, we find that the law of how this nonlinear effect is regulated by the intensity of the social dilemma in low-value games ($g_2$) is as follows: in scenarios with a strong social dilemma (small $r_2$), cooperative strategies are hard to sustain due to the conflict between individual rationality and collective interests. Increasing defection sensitivity further tightens the criteria for game state transitions, making the already fragile cooperative system less resistant to defection and raising the threshold $r_{2c}$ for sustaining cooperation. In contrast, in scenarios with a weak social dilemma (large $r_2$), cooperation already forms a preliminary group foundation. Increasing defection sensitivity at this stage incentivizes cooperative behavior, because cooperator clusters are more likely to switch to the high-value game $g_1$ and gain higher payoffs, thereby attracting more individuals to cooperate and leading to a sustained rise in the cooperation frequency $f_C$.

To further explore the nonlinear impact of defection sensitivity \(\alpha\) on \(f_C\) under higher-order interactions, we select three representative values of \(r_2\) for each system with \(g = 3\) and \(g = 4\), corresponding to three levels of social dilemma intensity in single low-value games \(g_2\): unsustainable cooperation, barely sustainable cooperation, and stably sustainable cooperation. Heatmaps are used to visualize the joint effects of defection sensitivity \(\alpha\) and the relative resource advantage of high-value games \(\delta = r_1 - r_2\) on \(f_C\), and the results are shown in Fig.~\ref{fig:exp2}.

In the strong social dilemma scenario shown in Fig.~\ref{fig:exp2}(\subref{fig:exp2_g3_collapse}) and (\subref{fig:exp2_g4_collapse}), the phase diagrams exhibit distinct pure defection regions. This indicates that the system\textquotesingle s cooperation frequency \(f_C\) depends entirely on the high-value game \(g_1\), and can only be sustained when the resource advantage \(\delta = r_1 - r_2\) is sufficiently large. As the defection sensitivity coefficient \(\alpha\), which quantifies how responsive hyperedges are to defection, increases, the boundary between the pure defection and mixed cooperation-defection regions shifts rightward, making it harder for cooperation to persist when \(g_1\) holds only a small resource advantage (small \(\delta\)).

\begin{figure*}[bp]
\centering
% First row
\begin{subfigure}[b]{0.32\textwidth}
\centering
\includegraphics[width=\textwidth]{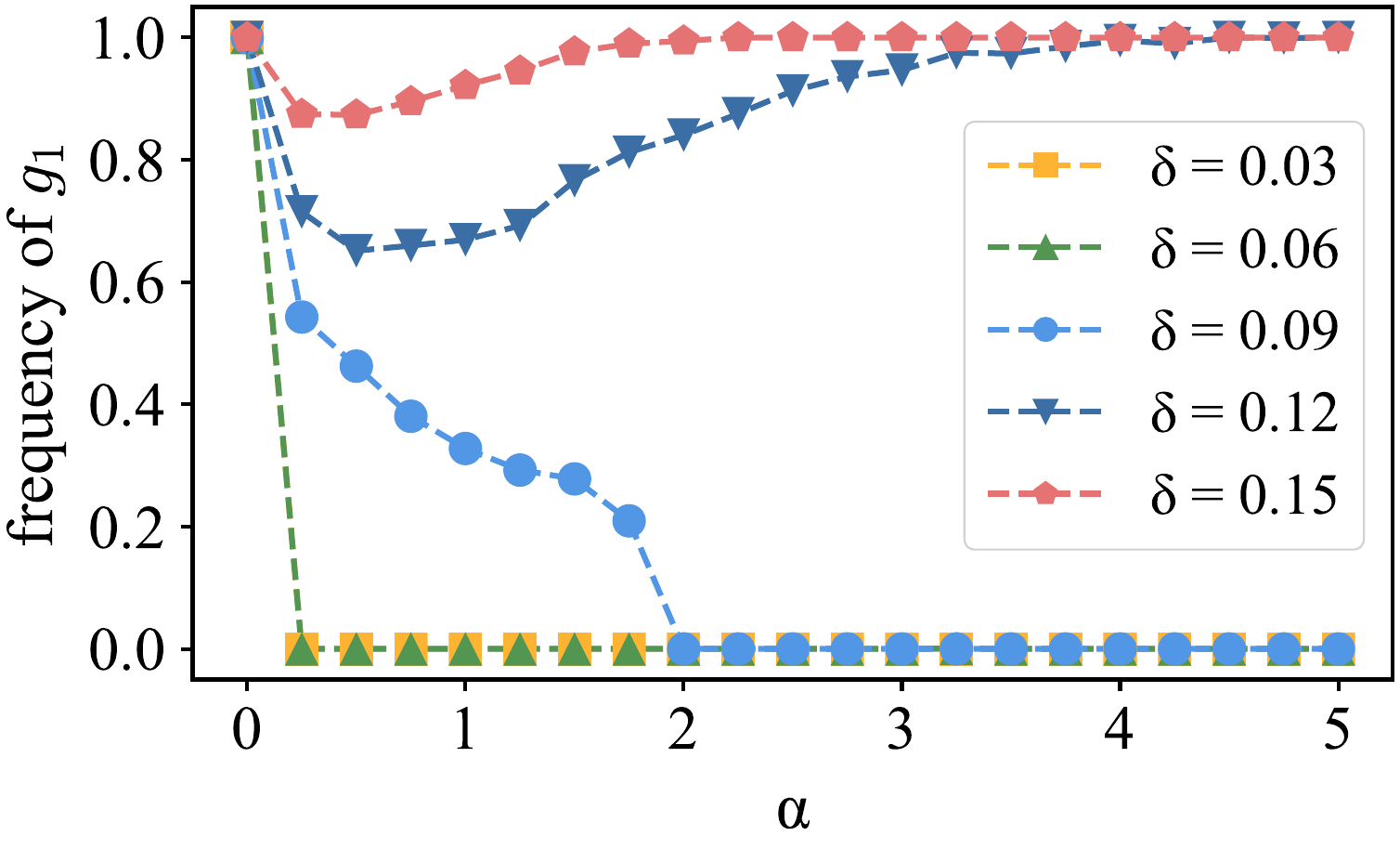}
\caption{\(f_{g_1}\): \(r_2=0.775\)}
\label{fig:exp5_f1_collapse}
\end{subfigure}
\hfill
\begin{subfigure}[b]{0.32\textwidth}
\centering
\includegraphics[width=\textwidth]{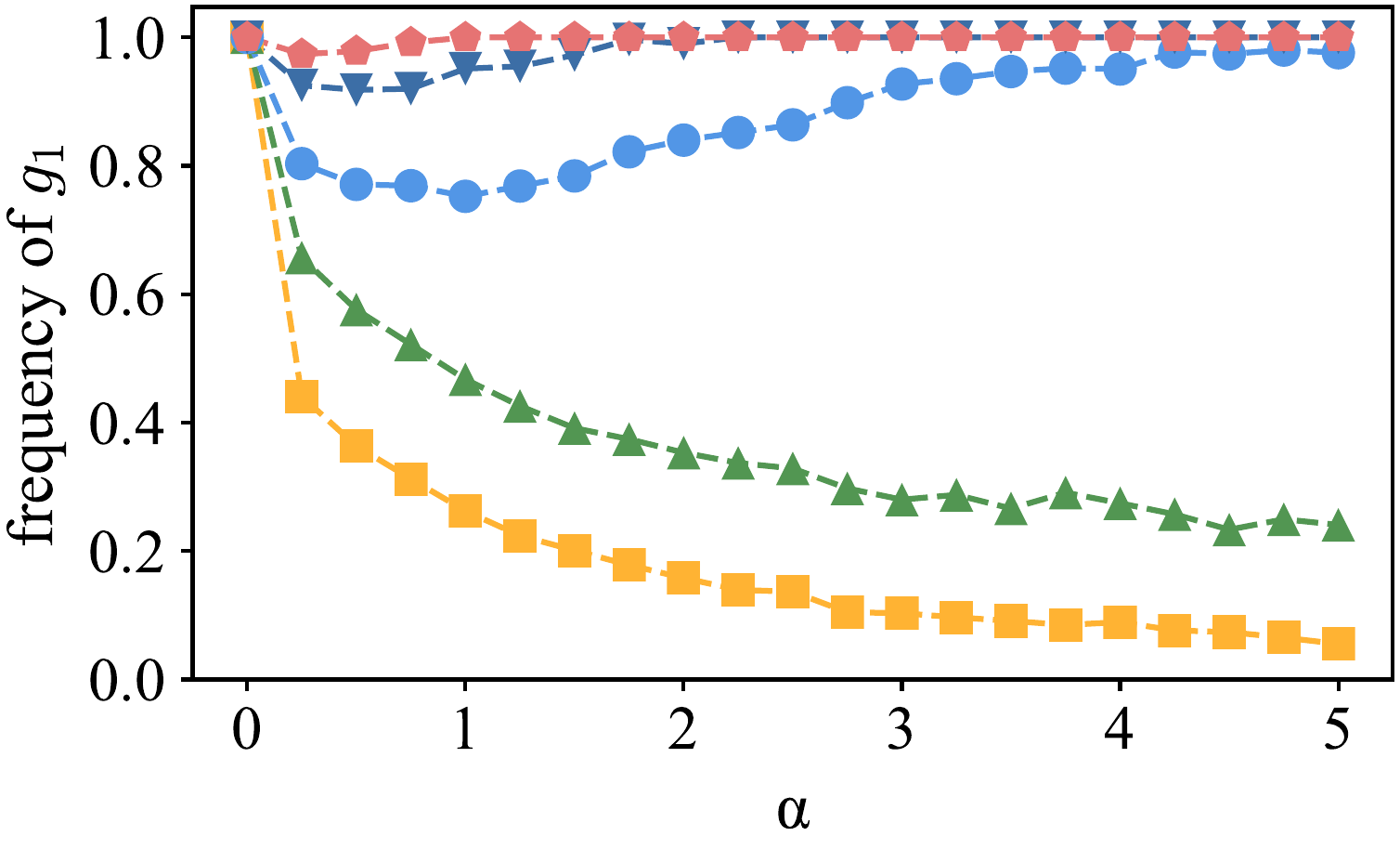}
\caption{\(f_{g_1}\): \(r_2=0.825\)}
\label{fig:exp5_f1_critical}
\end{subfigure}
\hfill
\begin{subfigure}[b]{0.32\textwidth}
\centering
\includegraphics[width=\textwidth]{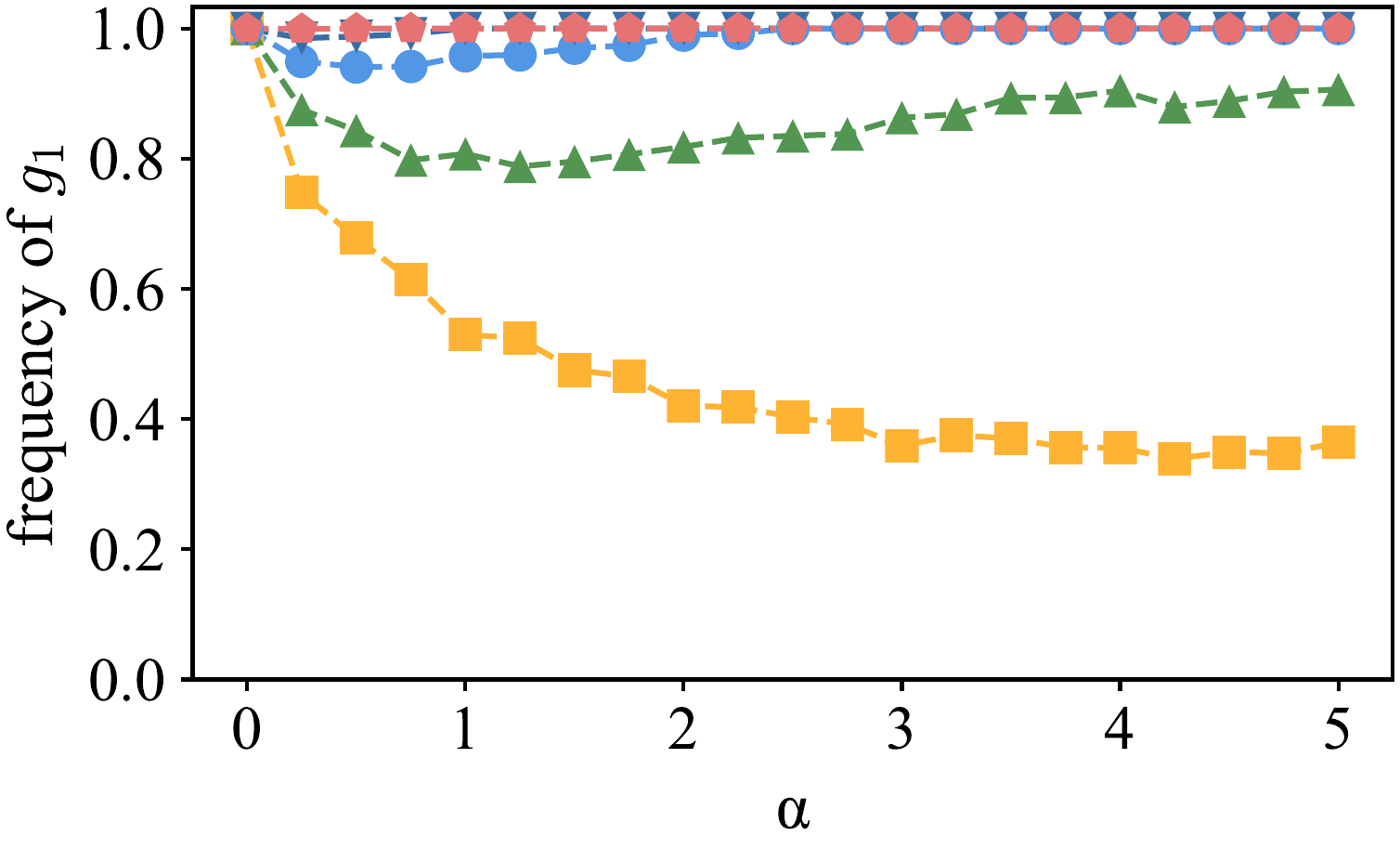}
\caption{\(f_{g_1}\): \(r_2=0.875\)}
\label{fig:exp5_f1_sustain}
\end{subfigure}

% Second row
\vspace{0.1cm}
\begin{subfigure}[b]{0.32\textwidth}
\centering
\includegraphics[width=\textwidth]{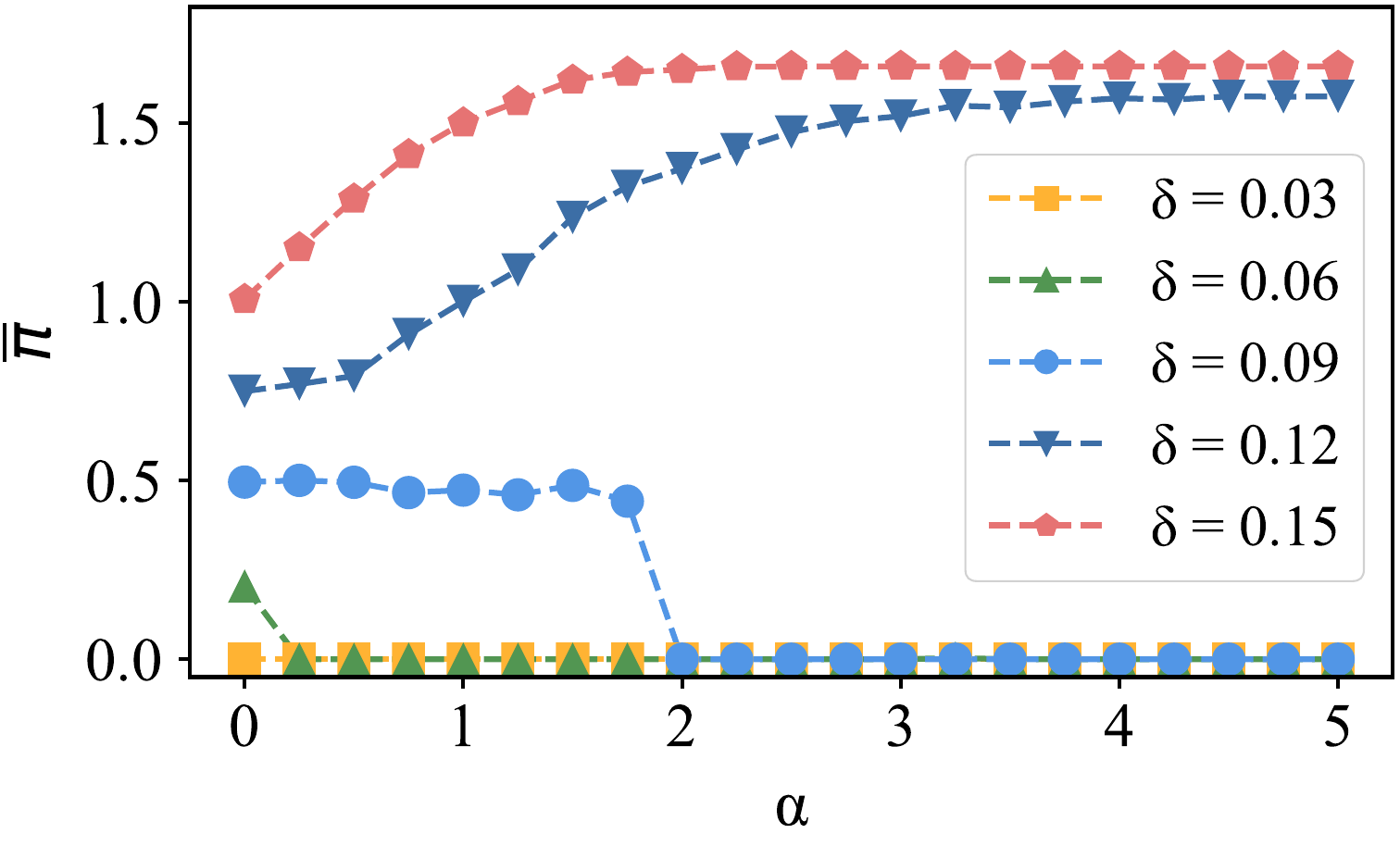}
\caption{\(\overline{\pi}\): \(r_2=0.775\)}
\label{fig:exp5_pay_collapse}
\end{subfigure}
\hfill
\begin{subfigure}[b]{0.32\textwidth}
\centering
\includegraphics[width=\textwidth]{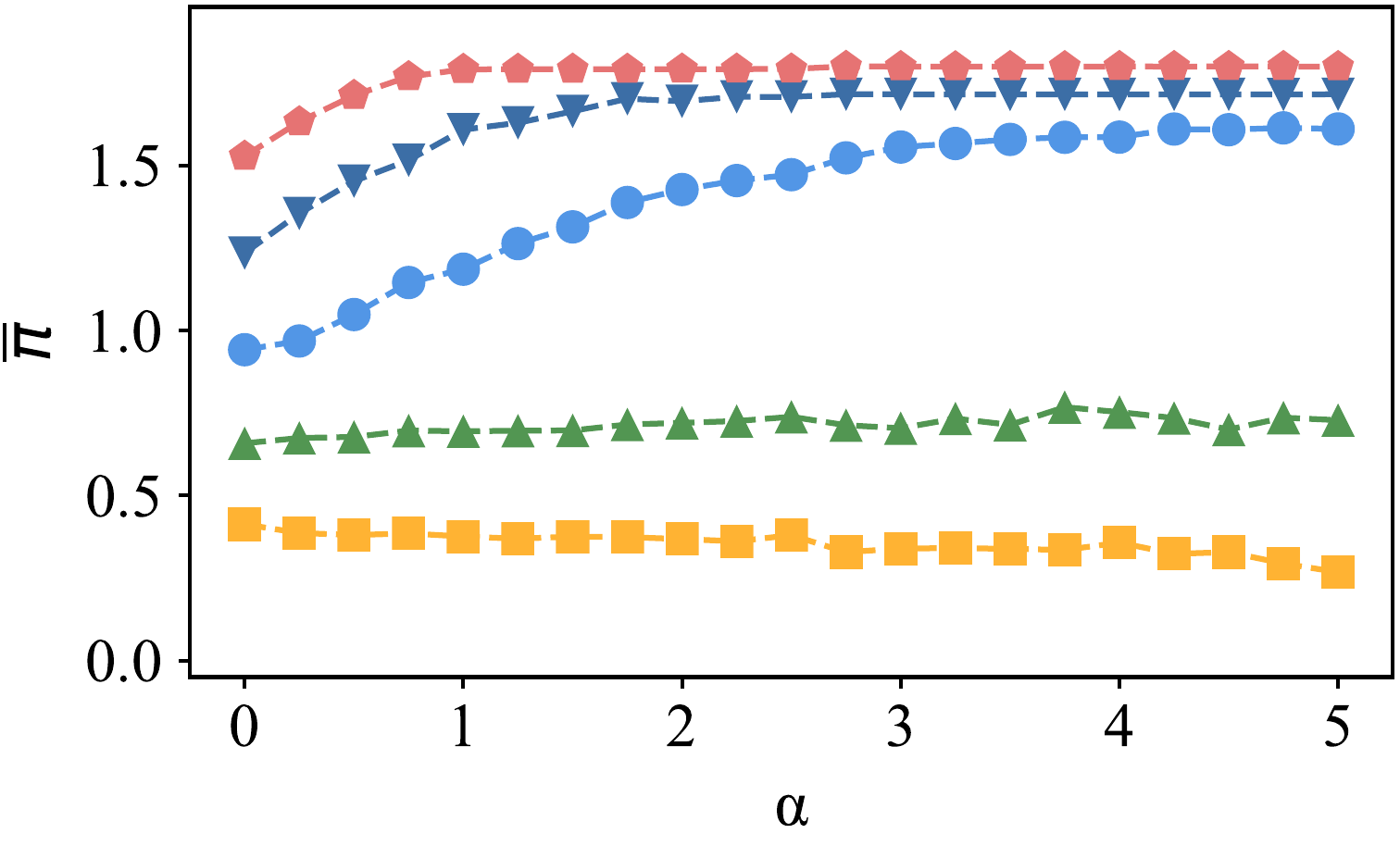}
\caption{\(\overline{\pi}\): \(r_2=0.825\)}
\label{fig:exp5_pay_critical}
\end{subfigure}
\hfill
\begin{subfigure}[b]{0.32\textwidth}
\centering
\includegraphics[width=\textwidth]{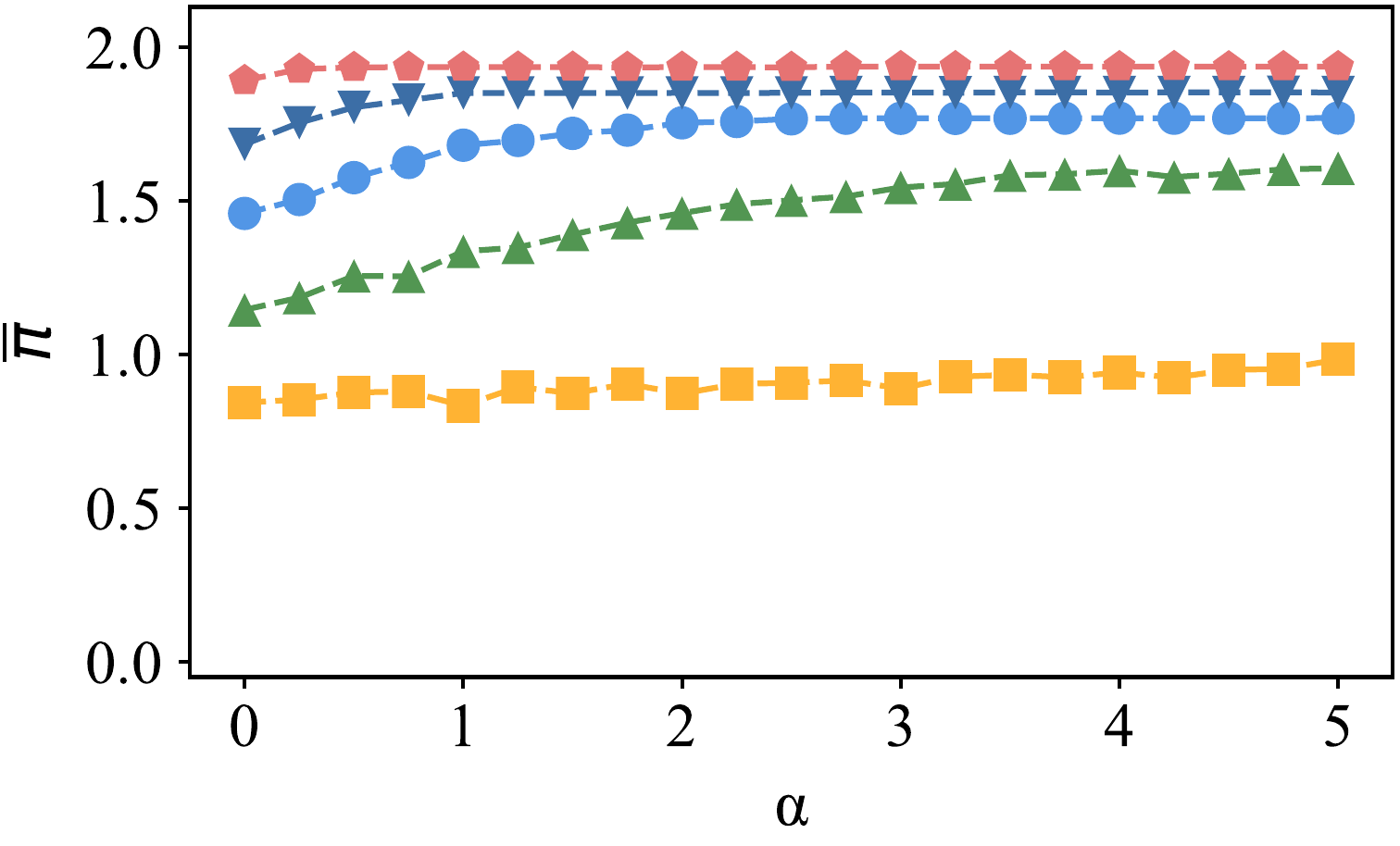}
\caption{\(\overline{\pi}\): \(r_2=0.875\)}
\label{fig:exp5_pay_sustain}
\end{subfigure}

\caption{\textbf{\(f_{g_1}\) and \(\overline{\pi}\) dependence of $\alpha$ for three representative \(r_2\) values.} \(g=3\) is fixed for all panels while the different values of \(\delta\) are indicated in the legend. Left, middle, and right columns show low, intermediate, and high $r_2$ values, which are 0.775, 0.825, and 0.875, respectively. In each column, panels exhibit cooperative positive feedback when \(\delta\) is large. For small \(\delta\), in the first column, \(g_2\) cannot sustain cooperation independently, causing \(f_{g_1}\) and \(\overline{\pi}\) to decline to 0.  In other panels, \(f_{g_1}\) decreases but remains non-zero, while \(\overline{\pi}\) stays stable.}
\label{fig:exp5}
\end{figure*}

In the moderate social dilemma scenario shown in Fig.~\ref{fig:exp2}(\subref{fig:exp2_g3_critical}) and (\subref{fig:exp2_g4_critical}), no pure defection regions are observed. This is because the low-value game \(g_2\) alone can barely sustain cooperation when there is no resource advantage (\(\delta = 0\)). A moderate increase in the defection sensitivity coefficient \(\alpha\) shifts the pure cooperation regions leftward, expanding the range of \(\delta\) values that support full cooperation.

In the weak social dilemma scenario shown in Fig.~\ref{fig:exp2}(\subref{fig:exp2_g3_sustain}) and (\subref{fig:exp2_g4_sustain}), this pattern is further validated, and an appropriate increase in the defection sensitivity coefficient \(\alpha\) directly boosts the system\textquotesingle s overall cooperation frequency \(f_C\).

%改
Overall, when the intensities of the social dilemma in both game states are adjusted simultaneously, the nonlinear influence of the defection sensitivity coefficient $\alpha$ on the cooperation frequency $f_C$ can be categorized into three types of cooperative dynamics. As $\alpha$ increases, $f_C$ exhibits full extinction (dropping to 0), stable maintenance (remaining nearly unchanged), or positive feedback growth (rising continuously).

%改
The specific cooperative dynamics depend on the ability of $g_2$ to sustain cooperation independently and the resource advantage $\delta$ of $g_1$. When the low-value game $g_2$ corresponds to a strong social dilemma, it cannot sustain cooperation on its own. A high defection sensitivity $\alpha$ suppresses cooperation unless the high-value game $g_1$ provides a large resource advantage $\delta$. When $g_2$ corresponds to a weak social dilemma, it can sustain cooperation independently.
As $\alpha$ increases, the region of full cooperation expands, even if the resource advantage $\delta$ of $g_1$ is small.

To investigate the cause of this nonlinearity, we define $L_{g_1}$ as the number of $g_1$ hyperedges, where $f_{g_1}=L_{g_1}/L_c$ represents the proportion of $g_1$ in the system. This proportion is crucial for cooperation maintenance and positive feedback. For $g=3$, we select the same three $r_2$ values as in Fig.~\ref{fig:exp2} to represent three levels of social dilemmas with different intensities in the low-value game. Under these three levels of social dilemma intensity, we separately explore how the relative resource advantage $\delta$ of the high-value game $g_1$ affects the variations in $f_{g_1}$ and the average payoff $\overline{\pi}$, with the results shown in Fig.~\ref{fig:exp5}.

\begin{figure*}[b]
    \centering

    \begin{subfigure}[b]{0.24\textwidth}
        \centering
        \includegraphics[width=\linewidth]{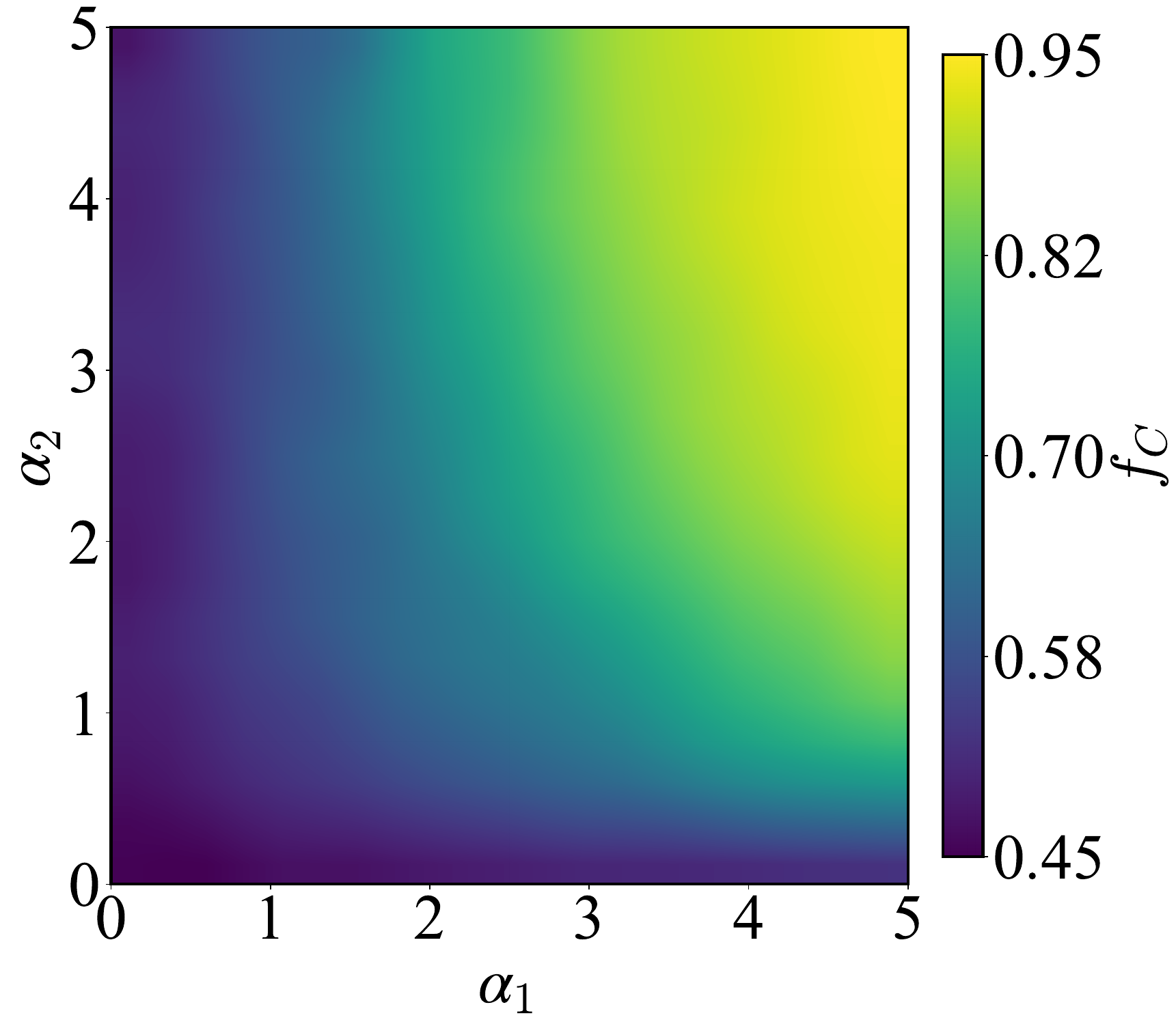}
        \caption{$f_C$, $\delta=0.09$}
        \label{fig:exp6_fc}
    \end{subfigure}
    \hfill 
    \begin{subfigure}[b]{0.24\textwidth}
        \centering
        \includegraphics[width=\linewidth]{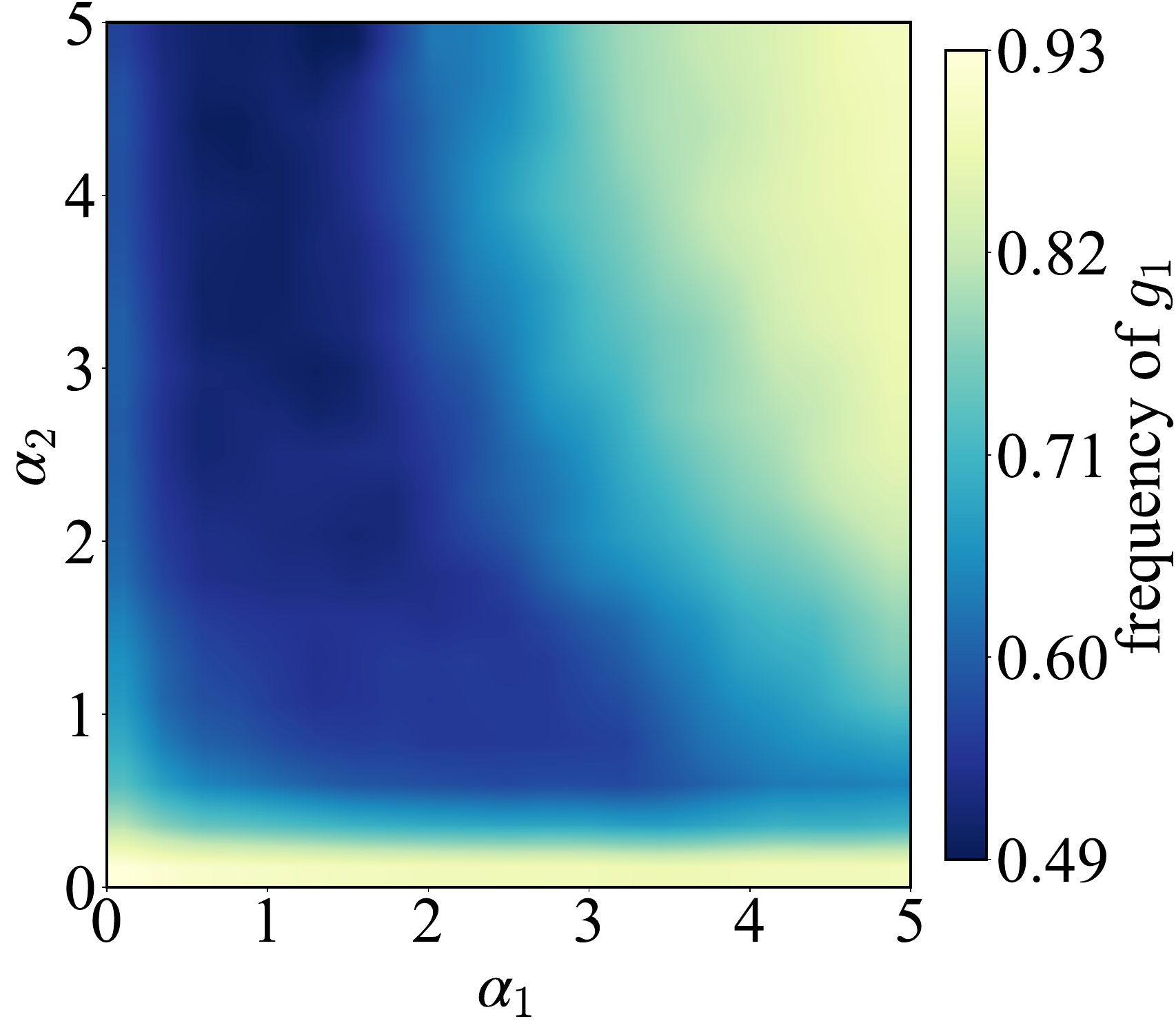}
        \caption{$f_{g_1}$, $\delta=0.09$}
        \label{fig:exp6_fg1}
    \end{subfigure}
    \hfill
    \begin{subfigure}[b]{0.24\textwidth}
        \centering
        \includegraphics[width=\linewidth]{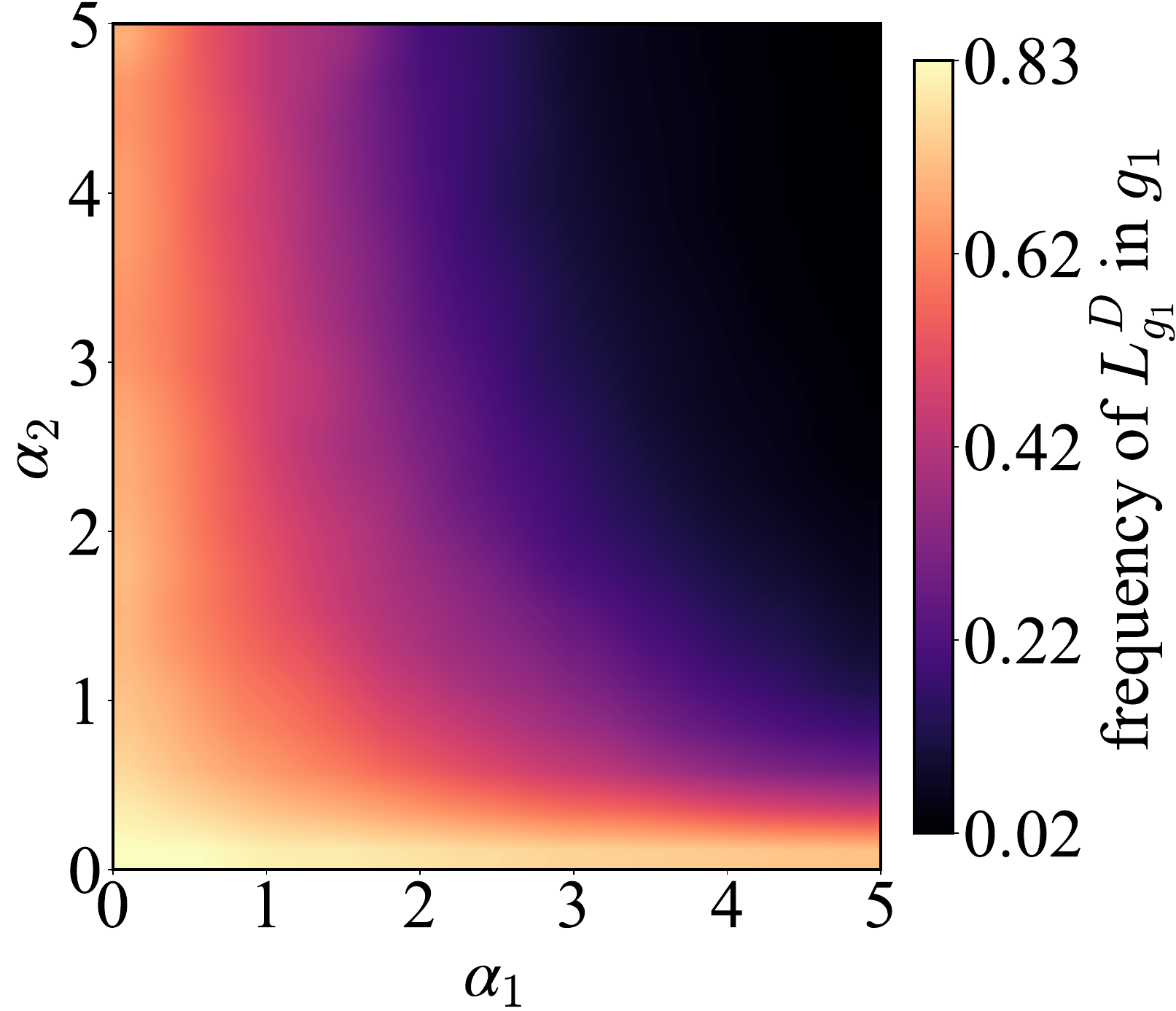}
        \caption{$f_{g_1}^D$, $\delta=0.09$}
        \label{fig:exp6_quality}
    \end{subfigure}
    \hfill
    \begin{subfigure}[b]{0.24\textwidth}
        \centering
        \includegraphics[width=\linewidth]{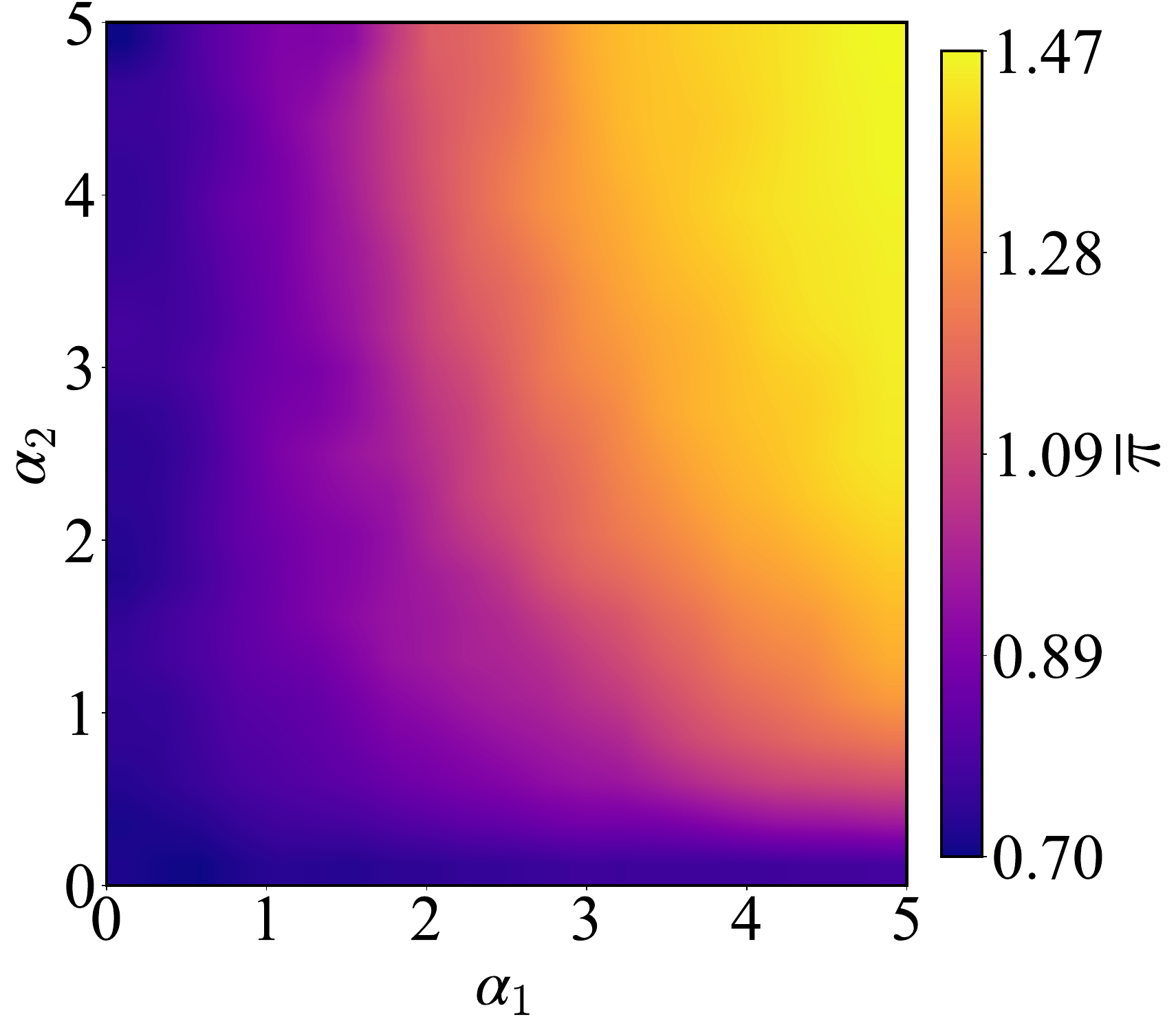}
        \caption{$\overline{\pi}$, $\delta=0.09$}
        \label{fig:exp6_pay}
    \end{subfigure}
    
    % 垂直间距调整，避免两行子图过密
    \vspace{0.1cm}
    
    % 第二行：对应原 fig:heatmaps_006（合作抑制场景，δ=0.06）
    \begin{subfigure}[b]{0.24\textwidth}
        \centering
        \includegraphics[width=\linewidth]{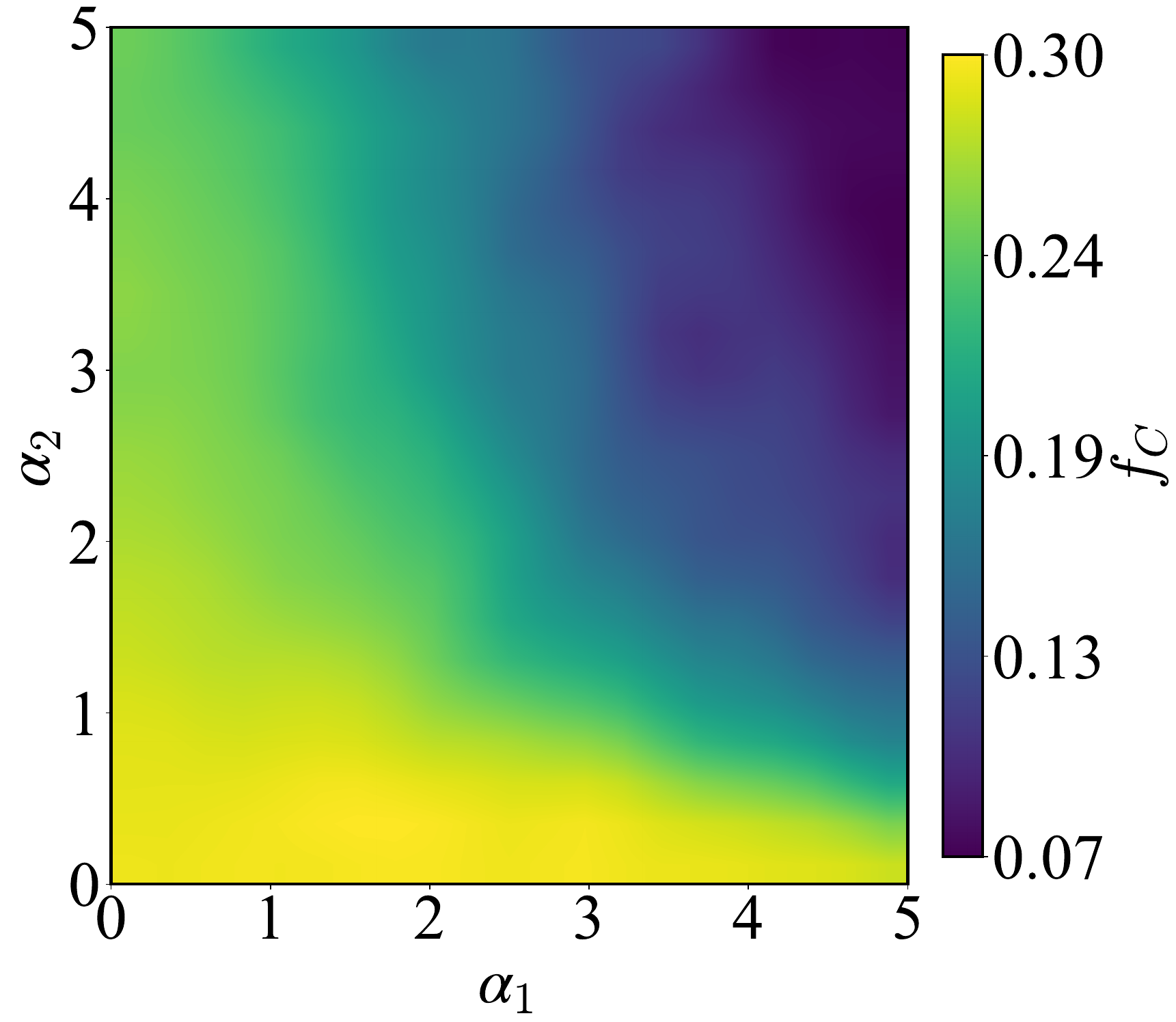}
        \caption{$f_C$, $\delta=0.06$}
        \label{fig:exp7_fc}
    \end{subfigure}
    \hfill
    \begin{subfigure}[b]{0.24\textwidth}
        \centering
        \includegraphics[width=\linewidth]{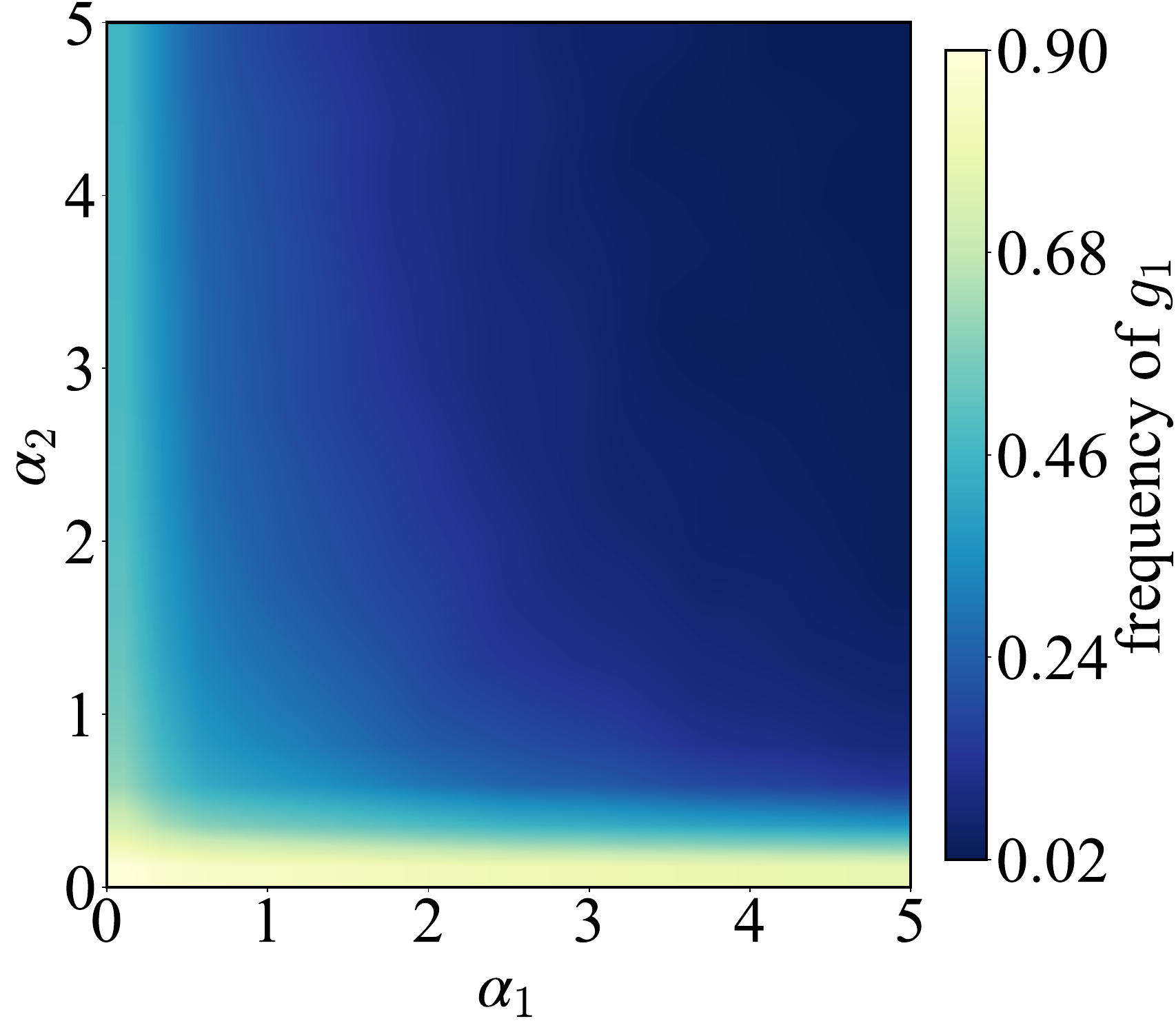}
        \caption{$f_{g_1}$, $\delta=0.06$}
        \label{fig:exp7_fg1}
    \end{subfigure}
    \hfill
    \begin{subfigure}[b]{0.24\textwidth}
        \centering
        \includegraphics[width=\linewidth]{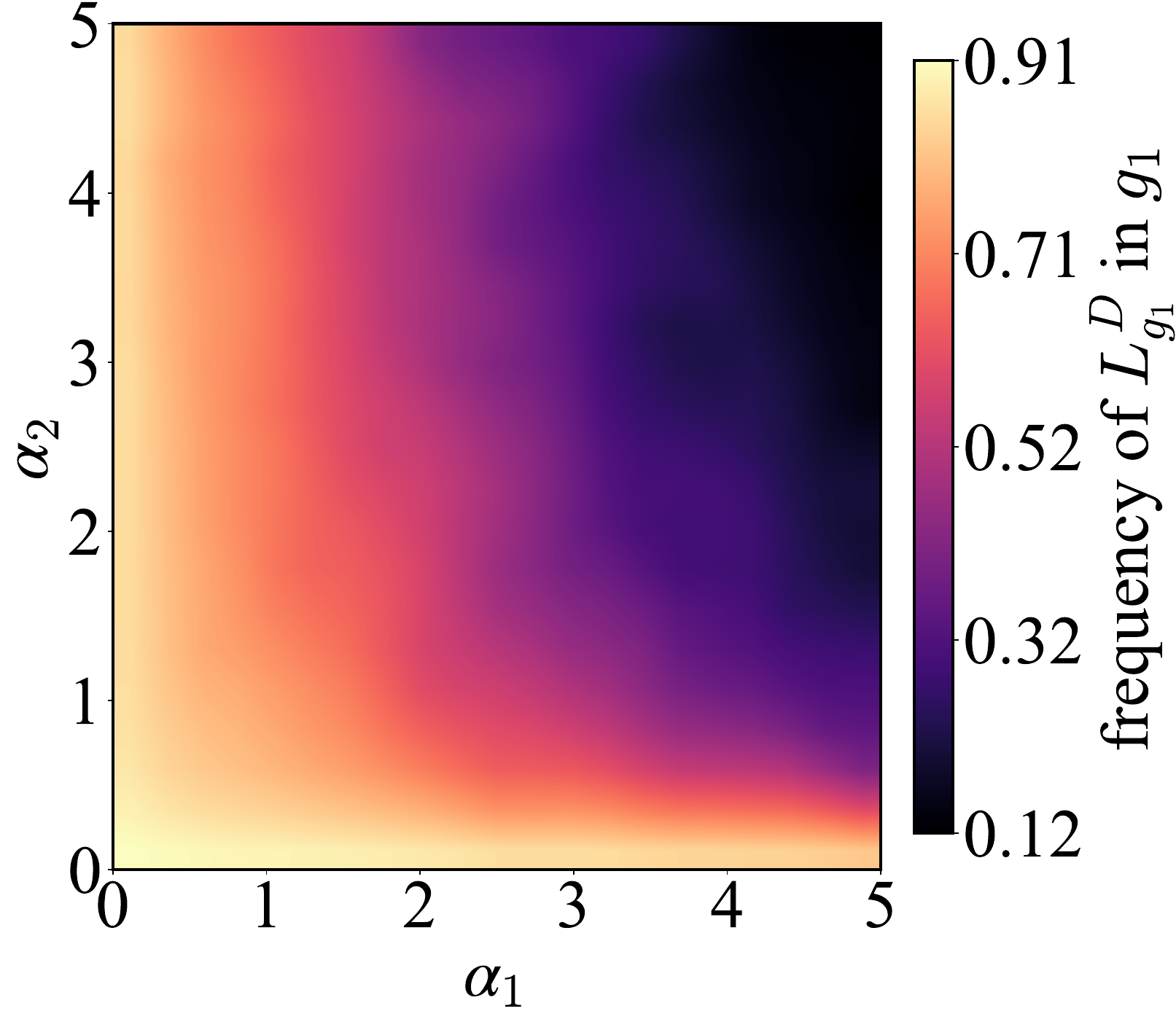}
        \caption{$f_{g_1}^D$, $\delta=0.06$}
        \label{fig:exp7_quality}
    \end{subfigure}
    \hfill
    \begin{subfigure}[b]{0.24\textwidth}
        \centering
        \includegraphics[width=\linewidth]{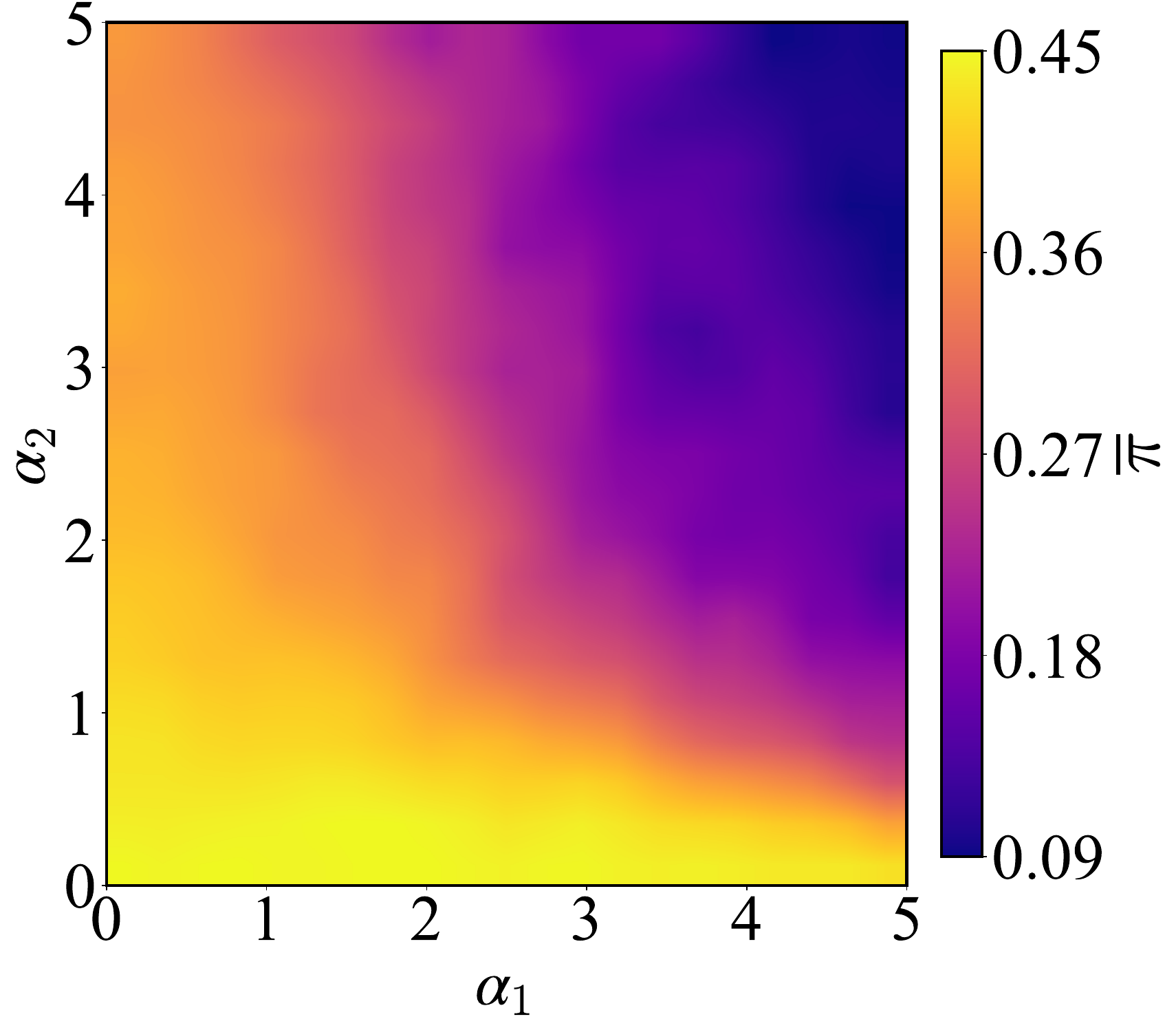}
        \caption{$\overline{\pi}$, $\delta=0.06$}
        \label{fig:exp7_pay}
    \end{subfigure}
    \caption{\textbf{Heatmaps of cooperation metrics under different \(\delta\) values.} Left to right in each row, panels show cooperation frequency \(f_C\), \(g_1\) proportion \(f_{g_1}\), defector-containing hyperedges \(f_{g_1}^D\), and average payoff \(\overline{\pi}\) respectively. Heatmaps are mildly smoothed with a Gaussian filter (\(\sigma=1\)) to reduce numerical noise, and bilinear interpolation is used for visualization clarity. The first row corresponds to the changes in various indicators when cooperation forms positive feedback, while the second row depicts the scenario where cooperation goes extinct as sensitivity increases. Results in the first column directly demonstrate the asymmetric variation of \(f_C\) with \(\alpha_1\) and \(\alpha_2\) under the state-dependent game transition scenario, and analyses of other indicators also exhibit a similar asymmetric phenomenon.}
    \label{fig:coop_metrics_heatmaps}
\end{figure*}

Fig.~\ref{fig:exp5} shows that positive feedback emerges across all $r_2$ scenarios, with the fraction of high-value games $f_{g_1}$ initially decreasing and then recovering. This occurs because a larger $\alpha$ enhances the cooperation purity of hyperedges in $g_1$. While $f_{g_1}$ declines, the cooperation quality of hyperedges within $g_1$ becomes higher. Cooperative clusters face a lower risk of exploitation by defectors and possess an intrinsic resource advantage, thereby gaining an increasingly strong payoff advantage. This drives a continuous rise in global cooperation, which further leads to the rebound of $f_{g_1}$.

The corresponding average payoff curve maintains an upward trend. This confirms that \(g_1\) can sustain a payoff advantage for cooperators. Concurrently, the social welfare of the system is reflected in the average payoff of agents \cite{Han2026Cooperation}. This upward payoff trend demonstrates that the promotion of cooperation is accompanied by a synchronous enhancement of social welfare. Notably, Fig.~\ref{fig:exp5}(\subref{fig:exp5_f1_collapse}) reveals that even when the low-value game \(g_2\) is in a strong social dilemma, the system can still trigger positive feedback if the resource advantage \(\delta\) of the high-value game \(g_1\) is significant enough to provide an adequate payoff advantage for cooperation.

Conversely, when \(\delta\) is small, the high-value game cannot sustain a payoff advantage for cooperation. The average payoff does not increase with the rise of defection sensitivity \(\alpha\), and the proportion of high-value games \(f_{g_1}\) decreases monotonically. At this point, the cooperative support capacity of the low-value game determines whether the system\textquotesingle s cooperation tends toward extinction or persistence. In Fig.~\ref{fig:exp5}(\subref{fig:exp5_f1_collapse}), when the low-value game \(g_2\) cannot sustain cooperation independently, the system evolves entirely into the low-value game as \(\alpha\) increases, ultimately leading to cooperation collapse. Correspondingly, Fig.~\ref{fig:exp5}(\subref{fig:exp5_pay_collapse}) demonstrates that social welfare decays to zero as cooperation collapses. In contrast, Fig.~\ref{fig:exp5}(\subref{fig:exp5_f1_critical}) and Fig.~\ref{fig:exp5}(\subref{fig:exp5_f1_sustain}) show that when the social dilemma of the low-value game \(g_2\) is alleviated and it possesses cooperative support capacity, naturally formed cooperator clusters can stabilize \(g_1\) at a non-zero level. At this point, social welfare also maintains a relatively stable trend. 

%改
This further explains how the two game states give rise to the three cooperative dynamics shown in Fig.~\ref{fig:exp2}.
Cooperation can form positive feedback only when the high-value game $g_1$ provides a significant resource and payoff advantage.
If $g_1$ fails to maintain such an advantage for cooperative behavior, the independent ability of the low-value game $g_2$ to sustain cooperation determines whether the system reaches stable cooperation or collapses into full defection. Furthermore, the numerical results demonstrate a notable positive correlation between cooperative behavior and social welfare under the game transition mechanism. The promotion of cooperative actions corresponds to enhanced social welfare, while their inhibition leads to reduced social welfare.

\subsection{State-Dependent Transitions}
Finally, we consider state-dependent transition probabilities, where hyperedges in different game states show varying sensitivities to defection by agents within the edges.

From previous analyses, when \( g_2 \) cannot sustain cooperation independently, the \( g_1 \) subsystem becomes critical for maintaining cooperation and even fostering positive feedback in the entire system under the transition mechanism. We characterize the importance of \( g_1 \) to system-wide cooperation using two metrics. The first is \( f_{g_1} \), the proportion of \( g_1 \) hyperedges in the entire system as previously defined. The second is \( f_{g_1}^D \), the fraction of hyperedges in \( g_1 \) that contain defectors (D). It is defined as
\begin{equation}
f_{g_1}^D = \frac{L_{g_1}^D}{L_{g_1}},
\label{fld}
\end{equation}
where \( L_{g_1}^D \) counts hyperedges with defectors in \( g_1 \), a lower \( f_{g_1}^D\) indicates higher levels of cooperative clustering and better hyperedge quality within \( g_1 \).

To investigate how asymmetric sensitivities affect agents' cooperative behaviors in the system, we use hypergraphs with \(g = 3\), fix the synergy factor of low-value games at \(r_2 = 0.8\), and set \(\delta = 0.06\) and \(\delta = 0.09\). In Fig.~\ref{fig:coop_metrics_heatmaps}, we illustrate the variations of various indicators with the two sensitivity factors under both cooperative positive feedback and cooperation collapse scenarios.

When \(\delta = 0.09\), Fig.~\ref{fig:coop_metrics_heatmaps}(\subref{fig:exp6_pay}) shows that players' average payoff undergoes a significant increase with the rise of the two sensitivities. That is, cooperative behavior can result in a payoff advantage, and a cooperation supporting feedback is formed. Fig.~\ref{fig:coop_metrics_heatmaps}(\subref{fig:exp6_fc}) reveals an asymmetry in how \(\alpha_1\) and \(\alpha_2\) affect the cooperation frequency: high-value regions in the heatmap are concentrated in the right half. This feature is more pronounced in Fig.~\ref{fig:coop_metrics_heatmaps}(\subref{fig:exp6_fg1}) and (\subref{fig:exp6_quality}). Notably, when \(\alpha_2 = 0\), the scale of high-value games remains large, but the quality of hyperedges in \(g_1\) is low. This is because when \(\alpha_2 = 0\), all hyperedges in \(g_2\) transition to high-value games with a probability of 1 in the next step, granting \(g_1\) an absolute advantage in scale while significantly reducing the quality of its hyperedges. When \(\alpha_2 \neq 0\), unfiltered transitions of hyperedges from \(g_2\) to \(g_1\) -- which would degrade hyperedge quality -- are avoided. On this basis, a higher \(\alpha_1\) ensures the prompt elimination of defectors in \(g_1\), resulting in the concentration of high-value regions in the right half.

When \( \delta = 0.06 \), the asymmetric inhibitory effect of sensitivity differences on cooperation is also observed in Fig.~\ref{fig:coop_metrics_heatmaps}(\subref{fig:exp7_fc}). Through Fig.~\ref{fig:coop_metrics_heatmaps}(\subref{fig:exp7_pay}), it can be known that \(g_1\) fails to maintain cooperators' payoff advantages at this point, with low-value regions concentrated in the right half. Although increased sensitivity effectively reduces the proportion of hyperedges containing defectors in \( g_1 \) and improves system quality, Fig.~\ref{fig:coop_metrics_heatmaps}(\subref{fig:exp7_quality}) shows that hyperedges containing defectors account for approximately 91\% of all hyperedges in \( g_1 \). Increased \( \alpha_1 \) leads to massive filtering of such hyperedges. Meanwhile, \( g_2 \) itself has a weak cooperative foundation, and increased \( \alpha_2 \) significantly hinders transitions into \( g_1 \), causing continuous shrinkage of \( g_1 \)\textquotesingle s scale. As shown in Fig.~\ref{fig:coop_metrics_heatmaps}(\subref{fig:exp7_fg1}), \( g_2 \) -- which cannot sustain cooperation independently -- dominates the system, leading to the collapse of cooperation. This further indicates that the imbalance between hyperedge quality and scale, caused by \( g_1 \)\textquotesingle s failure to maintain cooperators' payoff advantages, is the key reason for cooperation failure in state-dependent transitions.

Analysis of the two core cooperative dynamics -- positive feedback and cooperation collapse -- reveals that the two sensitivity parameters control how easily hyperedges switch to the high-value game \(g_1\), directly shape key cooperative outcomes, and play distinct roles in regulating \(g_1\)\textquotesingle s quality and scale: for quality control, \(\alpha_1\) filters out low-quality hyperedges to maintain the cooperative purity of original \(g_1\) hyperedges while \(\alpha_2\) ensures hyperedges transitioning from \(g_2\) to \(g_1\) meet quality standards and prevents unregulated quality decline; for scale control, \(\alpha_1\) limits the number of hyperedges leaving \(g_1\) by weeding out low-quality ones whereas \(\alpha_2\) governs the number of hyperedges switching from \(g_2\) to \(g_1\). While higher defection sensitivity improves \(g_1\) hyperedge quality (\(f_{g_1}^D\)), this alone cannot generate cooperative positive feedback -- the scale of \(g_1\) (\(f_{g_1}\)) is equally critical, as positive feedback only forms when \(g_1\)\textquotesingle s quality and scale improve in tandem; imbalances between these two indicators will cause continuous declines in \(f_{g_1}\) and eventually lead to cooperation collapse.

\section{Conclusion}
\label{sec:conclusion}
In this study, we propose and investigate a public goods game model by using game transition mechanisms based on URH, aimed at addressing the limitation of traditional pairwise interaction systems in capturing complex dynamics between individual strategies and group environments, and revealing the coevolution of individual strategies and group game states in collective interactions. 

We examine the coevolution of strategy selection and game state transitions in Public Goods Games on URH. The analysis of state-independent transitions reveals three characteristic cooperation patterns that emerge as the sensitivity parameter $\alpha$ increases: positive feedback, stable cooperation, and collapse of cooperation. The evolutionary outcome depends on the respective cooperation levels of the two game states. Extending the model to state-dependent transitions demonstrates asymmetric effects of the sensitivity parameters. Furthermore, we explain that the collapse of cooperation lies in the imbalance between the quality and scale of the high-value game system, which is induced by the system\textquotesingle s failure to maintain the payoff advantage of cooperation. 

%改
Future studies can address the limitations of the proposed model in three directions.
First, developing a theoretical framework to analyze the evolutionary dynamics of higher-order and pairwise interactions is a valuable research direction.
It will also help reveal the coevolution mechanisms between strategies and the environment from a theoretical perspective.
Second, we can incorporate agents' anticipatory behaviors toward game states.
This allows individuals to predict the environmental consequences of their strategies and extend myopic decisions to forward-looking ones.
The reinforcement learning framework introduced by \cite{pi2025TPAMI} can provide methodological support for this extension.
Third, environmental states in the real world often exhibit historical dependence.
Future states may not be determined solely by the current state.
Introducing memory effects in state transitions, as well as delayed or asynchronous switching,
can more realistically characterize the coevolution between individual behaviors and the environment.

\end{multicols}

\enlargethispage{20pt}

\ethics{This work did not require ethical approval from a human subject or animal welfare committee.}

\dataccess{
No experimental data were used in this work. The core simulation code for our model is openly available in a GitHub repository (\url{https://github.com/Nan2332/Code_for_Cooperation_in_Public_Goods_Games_over_Uniform_Random_Hypergraphs_with_Game_Transitions}).
}

\aucontribute{\\Nankun Wei: Writing  --  review \& editing, Writing  --  original draft, Methodology, Investigation, Formal analysis. \\
  Xiaojin Xiong: Writing  --  original draft, Methodology, Investigation, Formal analysis. \\
  Qin Li: Writing  --  review \& editing, Validation, Methodology, Formal analysis. \\
  Minyu Feng: Writing  --  review \& editing, Writing  --  original draft, Supervision, Methodology, Funding acquisition, Conceptualization. \\
  Attila Szolnoki: Writing  --  review \& editing, Funding acquisition, Formal analysis, Conceptualization.
}

\competing{We declare we have no competing interests.}

\funding{This work is supported
by the Chongqing Social Science Planning Project
under Grant NO. 2025NDQN41, Natural Science
Foundation of Chongqing under Grant
NO. CSTB2025YITP-QCRCX0007, and National
Research, Development and Innovation Office
(NKFIH) under Grant No. K142948.}

\ack{We thank the editors and reviewers for their advice and dedicated efforts.}

%%%%%%%%%% Insert bibliography here %%%%%%%%%%%%%%

% \bibliographystyle{unsrt}  
% \bibliography{newref}      

% Due to the outdated compiler version of the submission system and the inability to adjust the compilation sequence, 
% we have copied the locally generated bbl file below to ensure the normal display of references. 
% The original BibTeX compilation commands have been commented out accordingly.

\end{document}